\documentclass[prb,twocolumn]{revtex4-1} % style for Physical Review B and AJP are similar

\usepackage{comment}
\usepackage{amsmath} % need for subequations
\usepackage{amssymb}

\usepackage{graphicx} % for figures

\usepackage{hyperref}
\hypersetup{colorlinks=true,linkcolor=red,citecolor=blue,urlcolor=black,bookmarksdepth=subsubsection} %http://en.wikibooks.org/wiki/LaTeX/Hyperlinks#cite_note-1

\begin{document}

\title{Bubble optics}
\author{Markus Selmke}
\affiliation{*Fraunhofer Institute for Applied Optics and Precision Engineering IOF, Albert-Einstein-Str. 7, 07745 Jena and Germany}
\email{markus.selmke@gmx.de}
\homepage{http://photonicsdesign.jimdo.com}
\date{\today}

\begin{abstract}
Starting from a peculiar image observed below a bubble floating at a water-air interface, the article analyzes several optical properties of these special types of refracting objects (coined \textit{bubble axicons}). Using mainly geometrical optics, their relation to common axicons, the shadow sausage effect and elementary optical catastrophes (caustics) are discussed.
\end{abstract}

\maketitle 

\section{Motivation}
Menisci formed around small objects in water are a fascinating topic. Either the interactions they induce (cf.\ the "cheerios" effect \cite{Vella2005}; water striders walking on water \cite{Walker1983}), or the caustics they create when illuminated with light: the shadow-sausage effect observed for water-immersed sticks \cite{Adler1967,Walker1988,Lock2003} or the beautiful patterns decorating enlarged shadows below floating leafs are good examples of the latter \cite{Walker1988,Lock2015}. This article describes a peculiar lens formed by the meniscus around individual floating bubbles at an interface. Apart from the phenomenon being easily observable in many situations in our everyday lifes, the importance of bubble optics may also be appreciated due to the major role of bubbles on or within water for ocean science \cite{Czerski2017} and global climate science (''bright water'' \cite{Seitz2011}).

\begin{figure}[bt]
\centering
\fbox{\includegraphics[width=\linewidth]{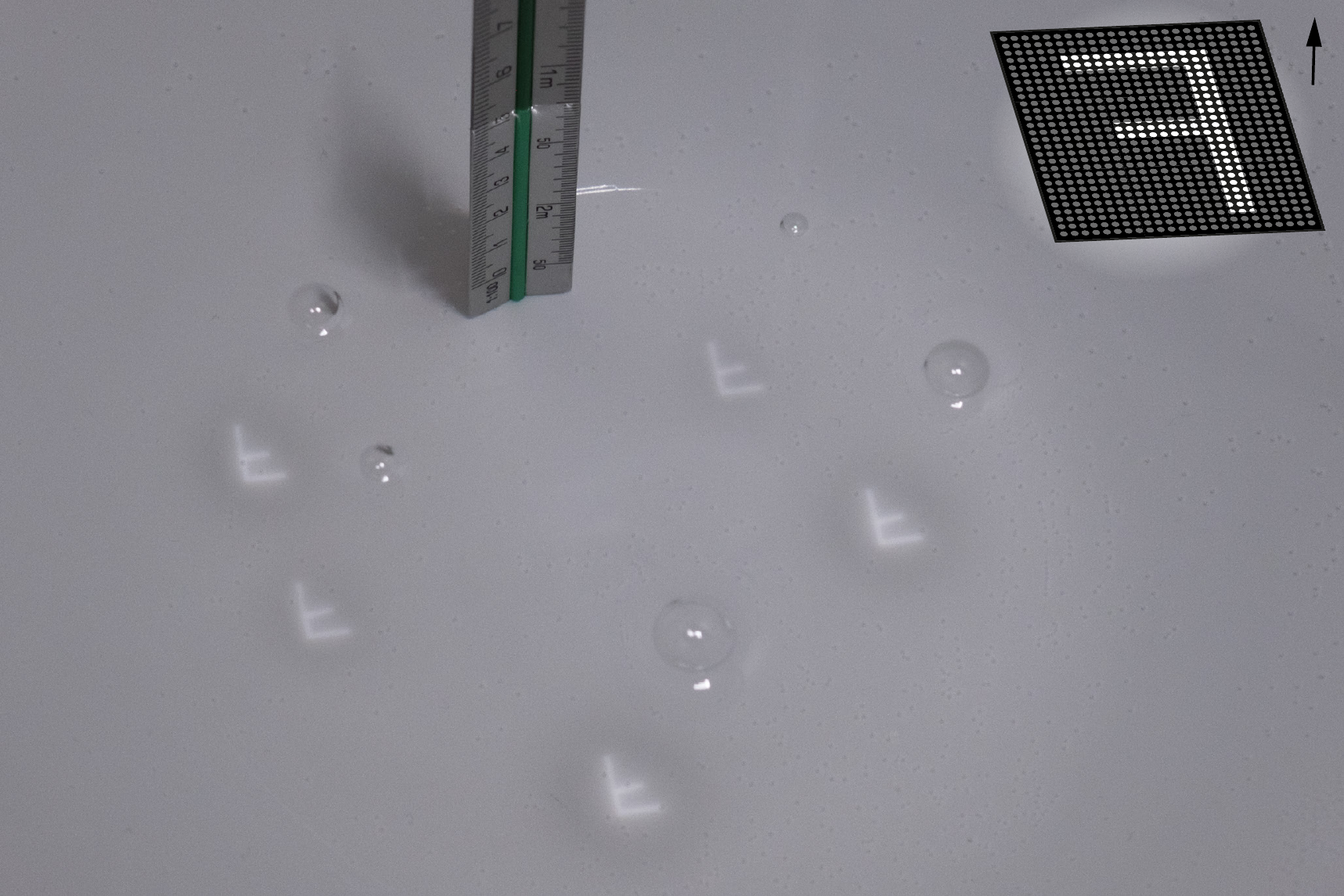}}
\caption{Surface bubbles act as special lenses to form same-sized images ${\sim 6.2\,\rm cm}$ below them at the bottom of a bowl. Here, an "F"-light source (ca.\ $13\,{\rm cm} \times 17\,\rm cm$) was realized by masking an LED matrix (Aputure Amaran HR672W) positioned ca.\ $1\,\rm m$ above the water level. Real inverted images appeared for virtually all bubble sizes for water depth $\gtrsim 3\,\rm cm$. The photo was contrast-enhanced to better show the dark halos surrounding the images. The effect is readily seen with the unaided eye.}
\label{Fig_Bath}
\end{figure}

The impetus for the present investigation was the author's surprise to see bright spots on the bottom of a bathtub while filling it. At this point, no soap was added to the water yet, and individual short-lived air bubbles (a strict two-phase water-air system) appeared sporadically: seemingly irrespective of the height of the water level, bright spots were clearly visible. Especially the latter observation seemed at odds with the usual behaviour of a lens having a limited focal range. While the bubbles were quickly identified to act as lenses producing images of the ceiling lights (recreated in a cereal bowl in Fig.\ \ref{Fig_Bath}), the large focal range was unexpected. The following investigations of the phenomenon's characteristics ensued. %Similar observations with wine glasses (which share some aspects in terms of the relevant geometry \cite{Selmke2018}) instead of bubbles have been reported \cite{Lunazzi2007}.

\begin{figure}[bt]
\centering
\fbox{\includegraphics[width=\linewidth]{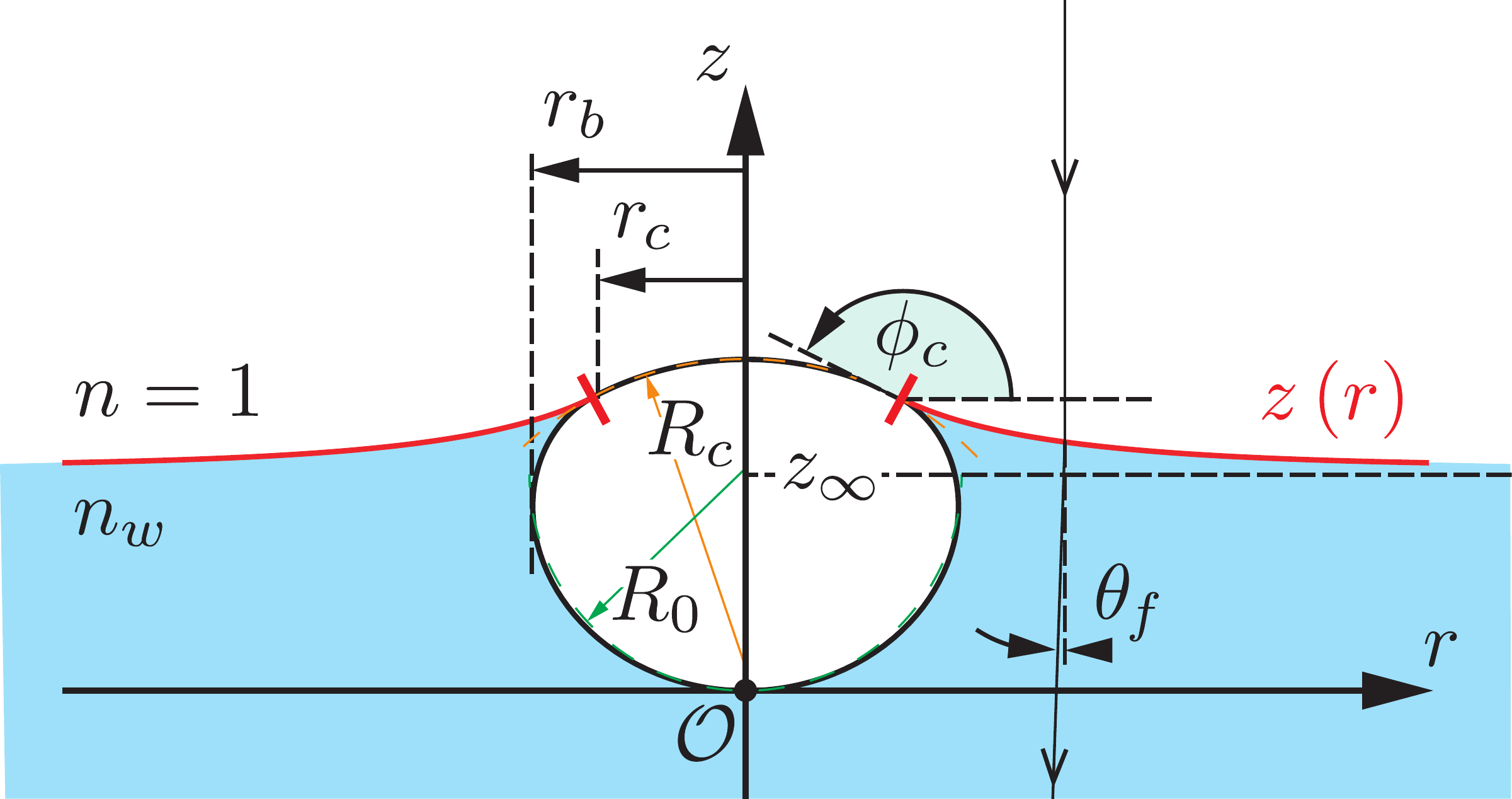}}
\caption{Sketch of the surface bubble geometry: The radius of curvature at the bottom of the bubble, which is the coordinate origin $\mathcal{O}$, is $R_0$, while the radius of the spherical cap (/dome) continuously connected to the upper part of the submerged bubble (cavity) and the outer meniscus is $R_c$ ("c" for "cap"). The bubble's maximum radius is $r_b$, while the rim of the outer meniscus (red) is at radial coordinate $r_c$. The sketch is for $R_0/a=1$, corresponding to a $\varnothing\equiv 2r_b \approx 5\, \rm mm$ diameter bubble on water ($n_w=1.33$).}
\label{Fig_Geom}
\end{figure}
Existing studies of light scattering for single bubbles have been limited to bubbles \textit{within} a fluid volume (cf.\ the works by Davis \cite{Davis1955} and Marston \cite{Marston1979}, and numerous more), or have dealt with foams \cite{Vera2001} or bubble rafts \cite{Dyson1949}. In a broader sense, the idea to optically interrogate menisci has been used in the past, for instance in interferometric studies of swimming particles \cite{Wardle1970,Hinsch1983}, analysis of their geometry by the image distortions they induce \cite{Mishra2015} or for nano particle detection \cite{DiLeonardo2003,Hennequin2013}. Also, refractometric studies have been done for swimming particles (which mostly \textit{depress} the water surface, especially regular floating objects), including their menisci \cite{Walker1988,Berry1983,Lock2015}.

However, although the latter two examples (and indeed a limiting case of the shadow-sausage effect \cite{Adler1967,Walker1988,Lock2003,Lock2015}) are closely related to this investigation, as will be explained later, no detailed study has yet been devoted to light refraction by single \textit{floating surface bubbles} (which \textit{raise} the water surface around them). Only brief notes or mentionings of the general phenomenon appear to exists in the literature \cite{Shields1990,Greenslade2012}, while single soap bubble optics focus on the colorful interference phenomenon instead.

In contrast, many experimental as well as theoretical studies exist on the \textit{shape} of a floating surface bubble \cite{BashforthAdams1883,Nicolson1949,Toba1959,Chappelear1961,Princen1963,Princen1965a,Medrow1971,Teixeira2015,Puthenveettil2018}, which for instance has been the starting point for studies on the fascinating dynamics and associated generation of aerosol particles in the collapse of bubbles (\cite{Lhuissier2012,Bird2010} and references therein). Likewise, it is be the the staring point for this investigation of their optical properties and is hence briefly recapitulated next.

\section{The shape of floating bubbles}\label{sec:bubbleshape}
The equilibrium shape of a floating surface bubble is the result of a tug of war between surface tension forces and pressures. It follows from the solution to the Young-Laplace equation (involving the principal radii of curvature $R_{1,2}$), which for the outer meniscus reads \cite{BashforthAdams1883,Nicolson1949,Toba1959,Chappelear1961,Princen1963,Princen1965a,Medrow1971,Teixeira2015,Puthenveettil2018}:
\begin{eqnarray}
\left(z\left(r\right)-z_\infty\right)\rho g&=&\sigma\left(\frac{1}{R_1}+\frac{1}{R_2}\right)\\
\frac{1}{R_1}+\frac{1}{R_2} &=& \frac{z''}{\left(1+z'^2\right)^{3/2}} + \frac{z'}{r\left(1+z'^2\right)^{1/2}}\nonumber\label{eq:Laplace},
\end{eqnarray}
with $z'=\mathrm{d}z/\mathrm{d}r$, $z''=\mathrm{d}^2z/\mathrm{d}r^2$, and appropriate boundary conditions: $z(r)\rightarrow z_\infty$ for $r\rightarrow \infty$, and $z'=\tan\left(\phi_c\right)$ at the junction $\left(r_c,z_c\right)$, see Fig.\ \ref{Fig_Geom}. To find the complete shape, three differential equations (cap / dome, submerged cavity / lower bubble surface, outer meniscus) need to be solved and matched together at the junction of these three domains. The natural length scale in this problem is the capillary length $a=\sqrt{\sigma/\rho g}$, where $g=9.81 \,\rm m/s^2$ is the gravitational acceleration, $\sigma$ is the surface tension of the liquid ($0.073 \,\rm N/m$ for water against air), and $\rho$ the density (density difference relative to air) of the liquid ($997\,\rm kg/m^3$ for water). For water, the \textit{capillary length} is about $a\sim 2.73\,\rm mm$, with all of the aforementioned parameters taken at room temperature and standard pressure. The Laplace pressure $\sim 2\sigma/R_0$, cf.\ Fig.\ \ref{Fig_Geom}, is considered to be negligible in its effect on the material properties such as $\sigma$ for the macroscopic bubbles treated here. Unfortunately, no analytical solution exists for the general case, and the numerical integration requires a shooting algorithm to guarantee that all three domains match and satisfy the boundary conditions. The reader is referred to the literature for details \cite{BashforthAdams1883,Nicolson1949,Toba1959,Chappelear1961,Princen1963,Princen1965a,Medrow1971,Teixeira2015,Puthenveettil2018}. A transition to gravity-dominated shapes for giant thin film bubbles only occurs at the meter scale \cite{Cohen2017}, which is not considered here.

Still, some qualitative remarks are in order to summarize the results of those previous shape investigations: Small bubbles are nearly spherical (with radius $R_0$) and almost completely submerged, with only a tiny cap above the water. Larger bubbles approach perfect hemispheres that float on top of the unperturbed water level and have a nearly flat interior cavity (bottom of the bubble). The cap of the bubble is always spherical, with a radius of curvature $R_c$. The exact shape in between these two limits is determined by the ratio of gravity $\rho g R_0$ (/hydrostatic pressure over a length $R_0$) to capillarity $\sigma/R_0$, i.e.\ by the Bond number $Bo=R_0^2/a^2$ ($R_0$ being the lower bubble radius of curvature, cf.\ Fig.\ \ref{Fig_Geom}). The quantity $R_0/a$ (or $Bo$) uniquely determines the shape of a bubble.
\begin{figure}[htbp]
\centering
\fbox{\includegraphics[width=\linewidth]{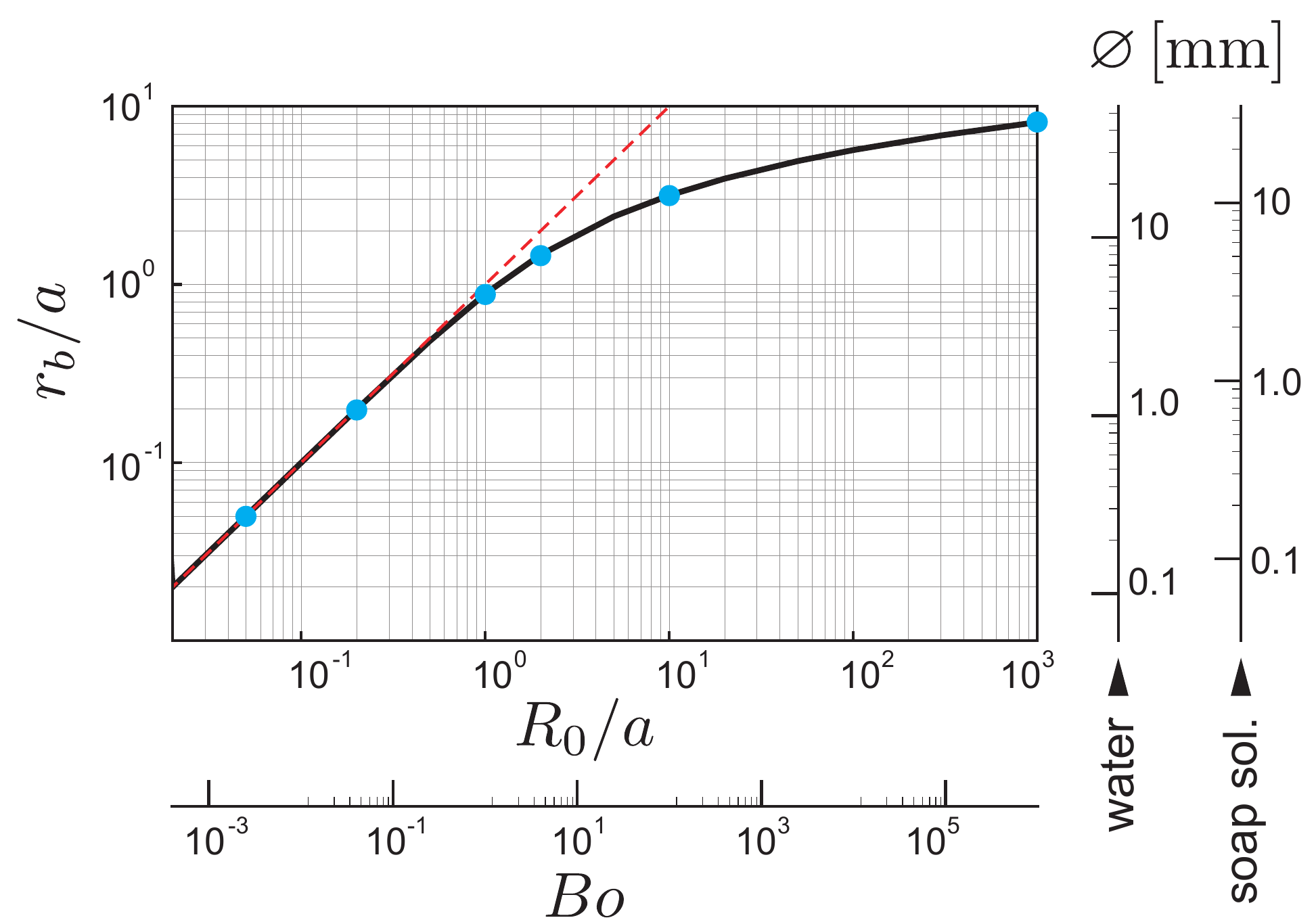}}
\caption{Relation between the bubble shape parameter $R_0/a$ (or $Bo$, bottom axes) and the bubble radius $r_b/a$ (left axis, or as an absolute diameter $\varnothing=2r_b$ for pure water and soap solution, right axes). The red dashed line shows the limiting behaviour of $r_b=R_0$ in the small bubble regime. The blue markers show the parameters considered in Figs.\ \ref{Fig_Raytracing} and \ref{Fig_Intensities}. The bubble radius for large $R_0/a\gg 1$ seems, without having any proof or derivation, to follow $r_b/a \sim \ln(R_0/a)$.}
\label{Fig_Graph}
\end{figure}
For small bubbles with $R_0/a\ll 1$ (or $Bo\ll 1$, surface tension dominates), the cap's radius of curvature is twice the radius of curvature of the bottom bubble interface, $R_c\rightarrow 2R_0$. \cite{Nicolson1949,Chappelear1961,Lhuissier2012,Puthenveettil2018} The radial extent of the small cap is $r_c \rightarrow 2 R_0 \sqrt{R_0^2/3a^2}$ \cite{Puthenveettil2018,Nicolson1949}. Still in the same limit, the radius of the bubble $r_b$ is essentially the submerged bubble's then constant radius of curvature, $r_b\approx R_0$, see Fig.\ \ref{Fig_Graph}. For large bubbles, with $R_0/a\gg 1$ (or $Bo\gg 1$), the spherical cap radius equals the bubble's size, that is $r_b\approx R_c$. The depth of the bottom of the bubble below the water surface at infinity, $z_\infty$, is given by $z_\infty / a^2 = 4/R_c - 2/R_0$, which means that $z_\infty / a \rightarrow 0$ holds for large bubbles where both $R_c/a \gg 1$ and $R_0/a \gg 1$. It is also worth emphasizing, that the outer meniscus is always raised above the unperturbed fluid level, $z(r)-z_\infty >0$, for all bubble sizes, a fact which is the basis for the lensing action discussed in the next section.

As a practical note, although somewhat detrimental to the ease of observation of individual bubbles (as opposed to foams), but beneficial for a controlled study otherwise: the stability (longevity) of floating bubbles can be increased significantly by adding a surfactant, \cite{Bird2010} which also reduces the surface tension ($\sigma=0.03 \,\rm N/m$ and $a=1.74\,\rm mm$ for a typical commercial soap solution \cite{Teixeira2015,Cohen2017}).

\section{Ray paths through a surface bubble}
Based on the observations stated in the introduction, the hypothesis of the investigation was that some part of the bubble acts as a lens to create the images.

\begin{figure}[htbp]
\centering
\fbox{\includegraphics[width=\linewidth]{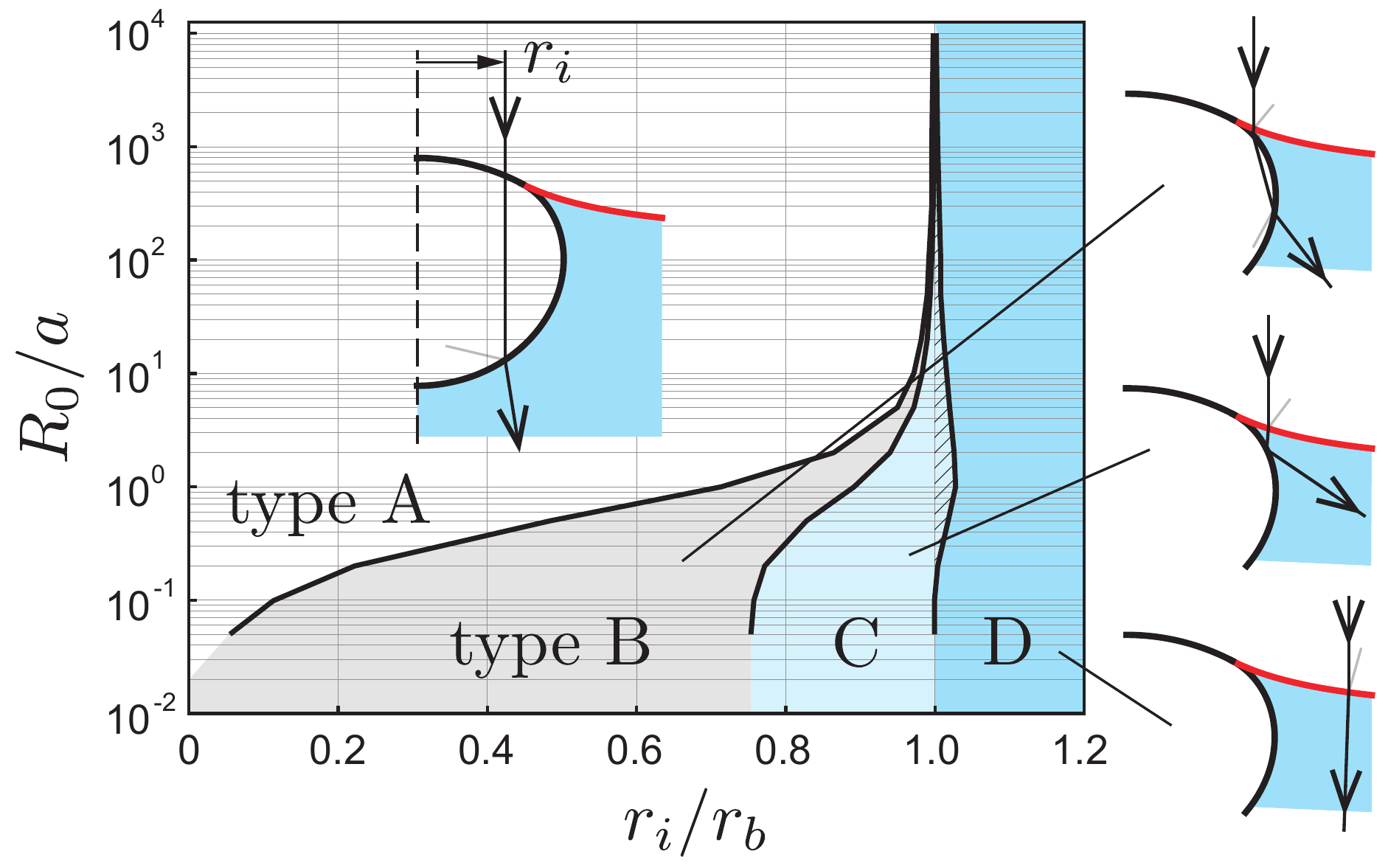}}
\caption{Phase diagram of the types of rays for different bubbles (parametrized by $R_0/a$, cf.\ this parameter's relation to $r_b/a$, in Fig.\ \ref{Fig_Graph}), depending on their impact parameter $r_i$ relative to the bubble's size $r_b$. [Note: The rightmost curve is basically the graph of Fig.\ \ref{Fig_fmin} (right axis) rotated by $90^{\circ}$ clockwise.]}
\label{Fig_Phase}
\end{figure}

\begin{figure*}[htbp]
\centering
\fbox{\includegraphics[width=\linewidth]{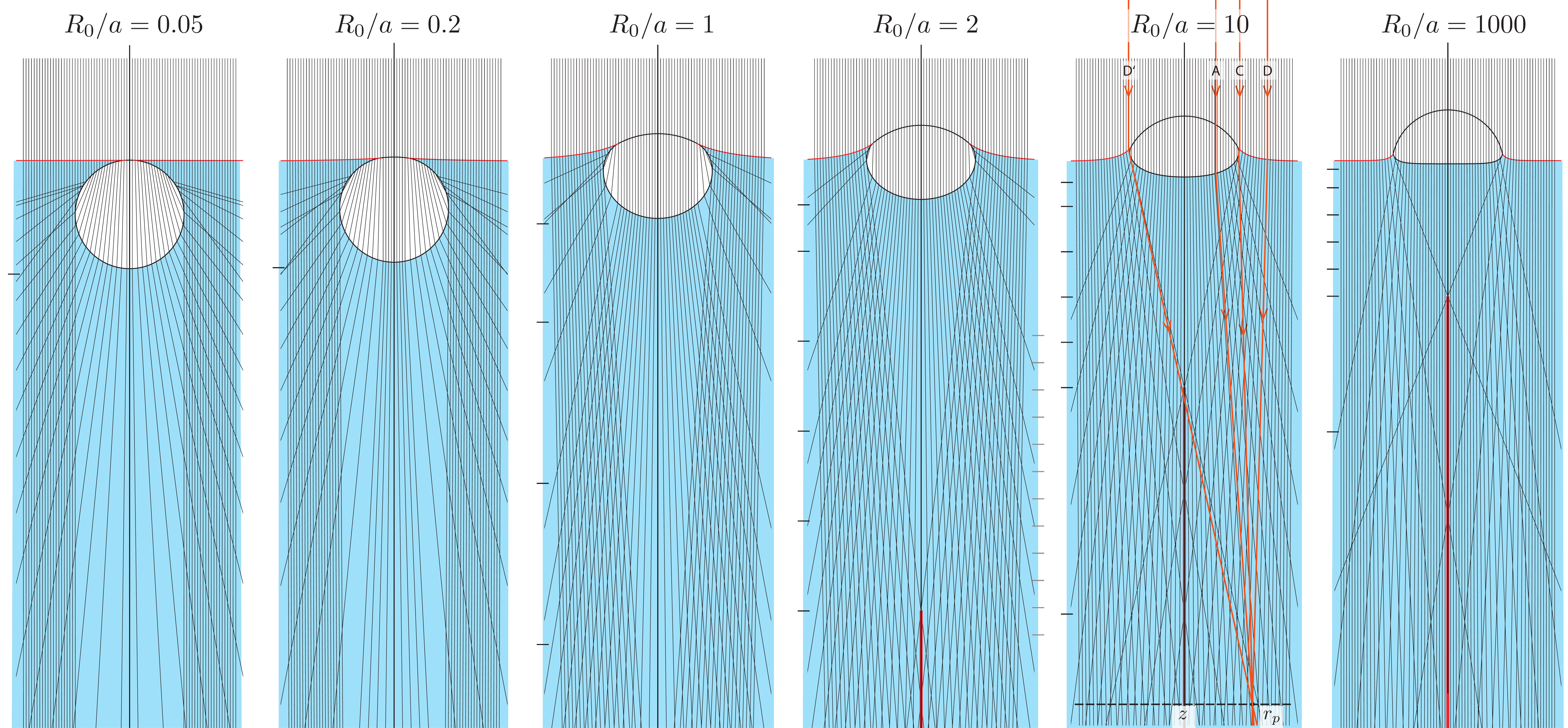}}
\caption{Ray tracing of several bubbles ($r_i=j\times r_{i,m}/20$, $j\in \mathbb{Z}$, $|r_i| < 2r_b$). The bubble shapes for different sizes of bubbles can also be found in loc. Fig.\ 2 of Toba's paper \cite{Toba1959}. The bubble sizes from left to right are $\varnothing=2r_b [\rm mm]=\left\{0.27, 1.1, 4.8, 8.0, 17, 44\right\}$, while the minimum focal distances are $f_m /a=\left\{122, 32.8, 13.1, 12.1, 13.2, 20.3\right\}$ and for water $f_m [{\rm cm}]=\left\{33, 8.9, 3.6, 3.3, 3.6, 5.5\right\}$. The orange rays added to the scenario of $R_0/a=10$ show the various ways (and ray types, cf.\ Fig.\ \ref{Fig_Phase}) in which light may contribute to a single point $(z, r_p)$ at a given screen distance $z_p$. The axial line caustic is due to types D rays from opposite sides (see thick red line for the $R_0/a\ge 2$-scenarios). The bold horizontal marks denote the depths $z_p$ for which Fig.\ \ref{Fig_Intensities} shows the intensity patterns, and the grey marks for $R_0/a=2$ those for Fig.\ \ref{Fig_Formation}.}%Fig_Geom \left\{124.1, 33.1, 13.15, 12.2, 13.4, 23.1\right\}$ $f_m [{\rm cm}]=\left\{34, 9.0, 3.6, 3.3, 3.7, 6.3\right\}$
\label{Fig_Raytracing}
\end{figure*}

First, bubble profiles parametrized by $R_0/a$ were calculated according to the exact theory using the shooting algorithm described in Appendix A of \cite{Lhuissier2012}. Using these profiles, a 2D-raytracer was written. In general, four major types of rays were identified, see type A - D in Fig.\ \ref{Fig_Phase} and the ray tracings in Fig.\ \ref{Fig_Raytracing}.

Type A rays enter (without appreciable refraction) with an impact parameter $r_i<r_c$ through the spherical thin film cap and get refracted at the lower air-water interface only. This lower bubble interface acts as a divergent lens, similar to a bubble in water. Accordingly, type A rays are always diverted and experience a deflection $\theta_f \approx \arcsin\left(r_i/R_0 \left[1/n_w - 1\right]\right) < 0$ away from the optical axis. 
%Plano-convex half-lens http://www.physicsinsights.org/simple_optics_spherical_lenses-1.html

Type B rays get refracted through the outer meniscus first, propagate through a portion of the outer meniscus before being refracted into the submerged bubble part, until finally being refracted out of it again. For air bubbles in water, these rays end up being always refracted away from the optical axes. For small bubbles, where $r_c\ll r_b\approx R_0$, the combined effect of this ray type together with type A is the appearance of a search-light beam emanating with some divergence from the bottom of the bubble, as can best be seen for $R_0/a=0.2$ ray tracings in Fig.\ \ref{Fig_Raytracing}.

Type C rays get refracted through the outer meniscus until encountering the submerged bubble at an angle to its normal larger than the angle of total internal reflection ($\theta_{\rm TIR}=\arcsin(1/n_w)\approx 48^{\circ}$). They hence get reflected externally from the submerged air-bubble and act to divert those incoming rays (mostly) away from the optical axes. Only within a very small parameter range (a subset of the dashed region in Fig.\ \ref{Fig_Phase}) will these rays maintain their meniscus-imparted convergence towards the optical axis via nearly tangential (/grazing) reflection from the bubble. If partial refraction according to the Fresnel equations is taken into account, type B and C may be subsumed under a single type. To clarify discussions, they have been labelled and are considered individually here.

Type D rays get refracted through the outer raised water meniscus only. These rays \textit{do get refracted towards the optical axes} and the outer meniscus hence acts as a converging lens (of sorts, see below). These rays may thus be seen to be the major contributors to the intensity distribution in the geometrical optics approximation far ($\gg r_b$) below illuminated bubbles. These rays are very much like the ones considered in previous studies on small floating particles \cite{Berry1983}, apart from the fact that they bend towards the axis which is uncommon for the inverse water depression menisci (and the associated enlarged shadows) typically formed around floating particles. For immersed vertical sticks, this crucial difference has been noted before by Walker \cite{Walker1988}, although without further analysis of the resulting focusing. Adler et al.\ \cite{Lock2015} provided some more details on this difference in the context of floating leafs, and the bubble phenomenon has been mentioned briefly in Refs.\ \cite{Shields1990,Greenslade2012}.

Other ray types, for instance those contributing to internal catacaustics or external reflextions have not been considered as they do not appear to add any major detail to the observations below a bubble described in this article.

\section{Approximation for small bubbles}
\subsection{Approximate shape of the outer meniscus}
An analytical approximation for the outer meniscus elevation $\Delta z$ above the unperturbed water level at large distances is available for small bubbles ($R_0/a\ll 1$) \cite{Nicolson1949,Puthenveettil2018}, \eqref{eq:zr}, where $\phi_c\lesssim 180^{\circ}$. The \textit{gentle slope approximation} of the Laplace equation of capillarity has also been explored for cylinders \cite{White1965,Huh1969,Tang2019}, spheres \cite{Berry1983} and more general geometries \cite{Hinsch1983,Lock2003}. It may be found from \eqref{eq:Laplace} by considering $(\mathrm{d}z/\mathrm{d}r)^2\ll 1$ and yields the profile:
\begin{equation}
\Delta z\left(r\right)=a C K_0\left(r/a\right), \quad C= \frac{r_c/a}{K_1\left(r_c/a\right) \sqrt{R_c^2/a^2 -r_c^2/a^2}}\label{eq:zr}.
\end{equation}
Herein, $K_0$ is the modified Bessel function of the second kind of order zero, and $C$ is constant for a given bubble. The approximation using $R_c$ \cite{Puthenveettil2018} instead of $R_0$ \cite{Nicolson1949} in the constant $C$ was found to give slightly better results, although for certain analytical simplifications in what follows the result $R_c\rightarrow 2R_0$ was used. The profile of the meniscus actually decays quickly as $\propto (r/a)^{-1/2}\exp(-r/a)$ for large $r/a\gg 1$ \cite{Lock2003}, with an amplitude $h=a C\rightarrow R_0 \times 2R_0^2/3a^2$.

\begin{figure}[htbp]
\centering
\fbox{\includegraphics[width=1\linewidth]{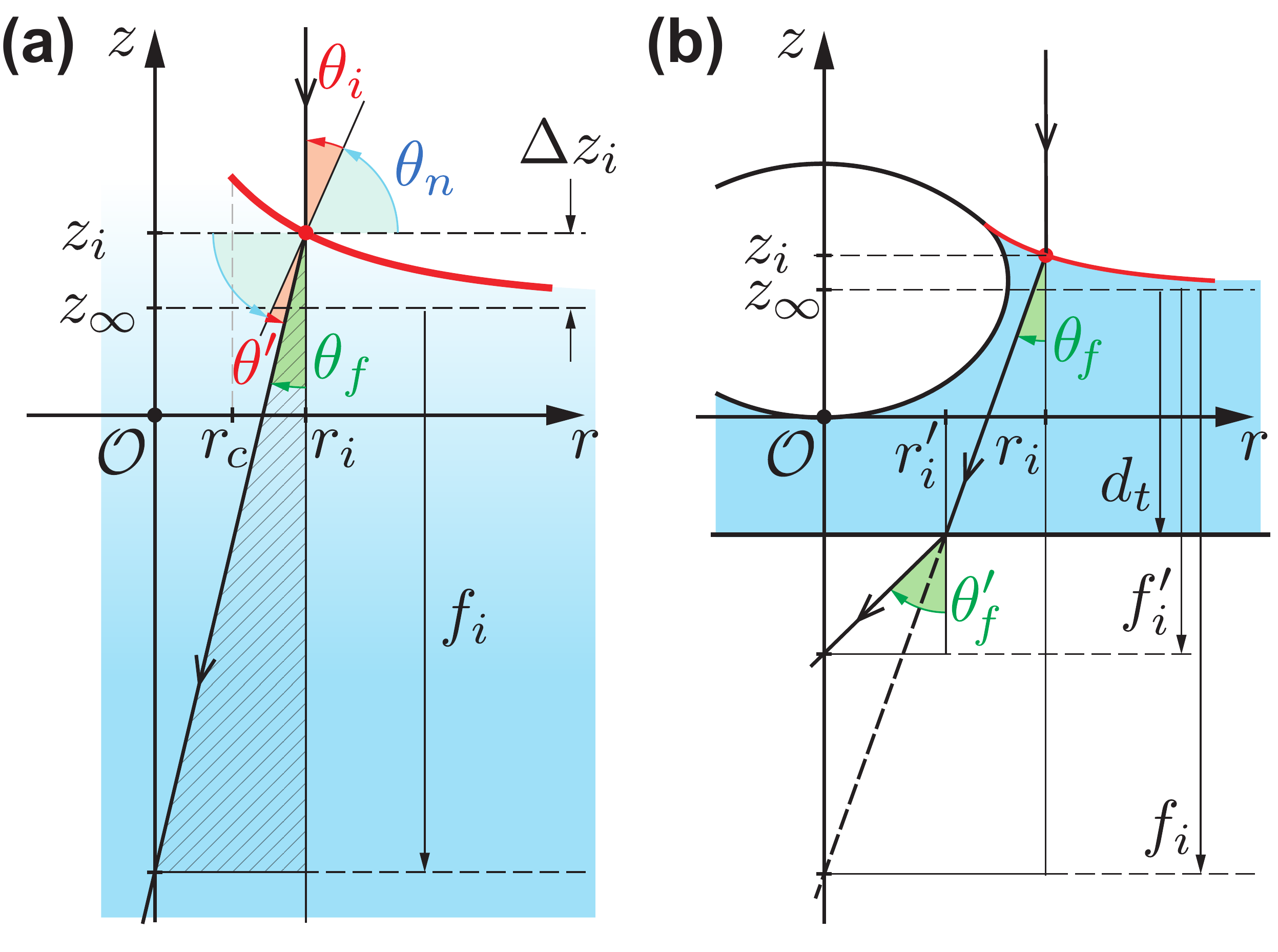}}
\caption{\textbf{(a)} Sketch of the geometry for the determination of the intercept depth $f_i=f\left(r_i\right)$ below the unperturbed water level at $z_\infty$. \eqref{eq:f} follows from considering the textured triangle. \textbf{(b)} Sketch of a convenient experimental setup (cf.\ also section \ref{sec:oblique}\ref{sec:observations}): a transparent water tank with a flat base. Using Snell's law, $\sin\theta_f'\approx n_w \sin \theta_f$, one finds from the sketch (and using some trigonometric relations): $f_i' \approx d_t+(f_i-d_t)n_w^{-1}{(1-(n_w^2-1)r_i^2/f_i^2)^{1/2}}$. That is, for a shallow tank with $(d_t,r_i) \ll f_i$ the intercepts and focal lengths are scaled approximately via the refractive index, $f_i' \approx f_i/n_w$.}%textured triangle, which has its apex at the entry point $(r_i,z_\infty + \Delta z(r_i))$ and an angle $\theta_f=\pi/2-\theta_n-\theta'$. 
\label{Fig_f}
\end{figure}

\subsection{Refraction geometry for type D-rays}\label{sec:Refraction}
Referring to Fig.\ \ref{Fig_f}(a), we can infer the angle $\theta_n$ of the surface normal from \eqref{eq:zr} via $\tan\left(\theta_n\right)=-1/(\mathrm{d}z/\mathrm{d}r)=1/C K_1\left(r/a\right)$, with the function evaluated at the vertical incident ray's impact parameter $r=r_i$. The angle of incidence of the ray to the meniscus is then $\theta_i=\pi/2 - \theta_n$, and from Snell's law of refraction the refracted ray's inclination towards the meniscus normal after refraction becomes $\theta'=\arcsin\left(\sin\left(\theta_i\right)/n_w\right)$. If we define the distance of the intercept with the optical axis below the surface level at infinity as $f_i$, and $\Delta z_i = \Delta z\left(r_i\right)$, one finds from Fig.\ \ref{Fig_f}
\begin{equation}
\tan\left(\pi/2-\theta_n - \theta'\right) = r_i/\left(f_i+\Delta z_i\right) \label{eq:f}.
\end{equation}
Using Taylor series expansions for expressions involving the small angles $\theta'$, $\theta_i$ and $\theta_f$, ignoring the small quantity $\Delta z_i$, and finally using the expression for $C$ in \eqref{eq:zr}, one finds
\begin{equation}
\frac{f_i(r_i)}{a} \approx \frac{r_i}{r_c}\frac{\sqrt{R_c^2/a^2 -r_c^2/a^2}}{\left(1-1/n_w\right)} \frac{K_1\left(r_c/a\right)}{K_1\left(r_i/a\right)} \label{eq:fapprox}.
\end{equation}
\eqref{eq:fapprox} can be used to show that the (negative spherical) aberrations are such that the focal length diverges as $f_i \propto \left(r_i/a\right)^{3/2} \exp\left(r_i/a\right)$, meaning that there is no single well-defined focal length of the bubble lens. Similar mappings of the radius of the annular region to a certain $z$-coordinate (here: $f_i(r_i)$) were also given for menisci around cylinders and floating objects in the gentle-slope approximation in the works of Lock et al.\ \cite{Lock2003} and Adler et al.\ \cite{Lock2015}. %loc.\ Eq.\ (18) of Ref.\ \cite{Lock2003}, and Adler et al.\ in Ref. [\citenum{Lock2015}].%, loc.\ Eq.\ (10).

\subsection{Approximation of the minimum focal distance}
Using further the approximation for small bubbles of $r_c$ \cite{Puthenveettil2018,Nicolson1949} stated in section \ref{sec:bubbleshape}, approximating the relevant impact parameter for the minimal focal distance ${min}\left(f_i\right)\equiv f_m=f\left(r_{i,m}=r_b\approx R_0\right)$ (see Fig.\ \ref{Fig_SketchApprox}(a)), and Taylor-expanding the remaining Bessel functions in \eqref{eq:fapprox} for small arguments $R_0/a$, one finally arrives at
%(for water, the actual minimum focus impact parameter $r_i$ was found to be $<3\%$ larger than $r_b$ for $R_0/a \le 1$, and $r_b/R_0 < 12\%$)
\begin{equation}
\frac{f_m}{a} \approx \frac{3}{2\left(1-1/n_w\right)} \left(\frac{R_0}{a}\right)^{-1}, \quad \frac{f_m}{R_0} \propto \left(\frac{R_0}{a}\right)^{-2}\label{eq:fapprox2}.
\end{equation}
\eqref{eq:fapprox2} shows that both on an absolute scale (fixed $a$) as well as on a relative scale (normalized to the bubble radius $R_0$) the focus shifts away from the surface with decreasing bubble size (where $r_b \rightarrow R_0$) in this limit. The latter is consistent with the diverging behaviour of the starting point of the focal region observed in the ray tracings of Fig.\ \ref{Fig_Raytracing}: Note how the focus shifts away from the bubble for the smaller bubbles of the series on the left. 

\section{Minimum focal distance: the general case}\label{sec:minfocdistance}
\begin{figure}[htbp]
\centering
\fbox{\includegraphics[width=\linewidth]{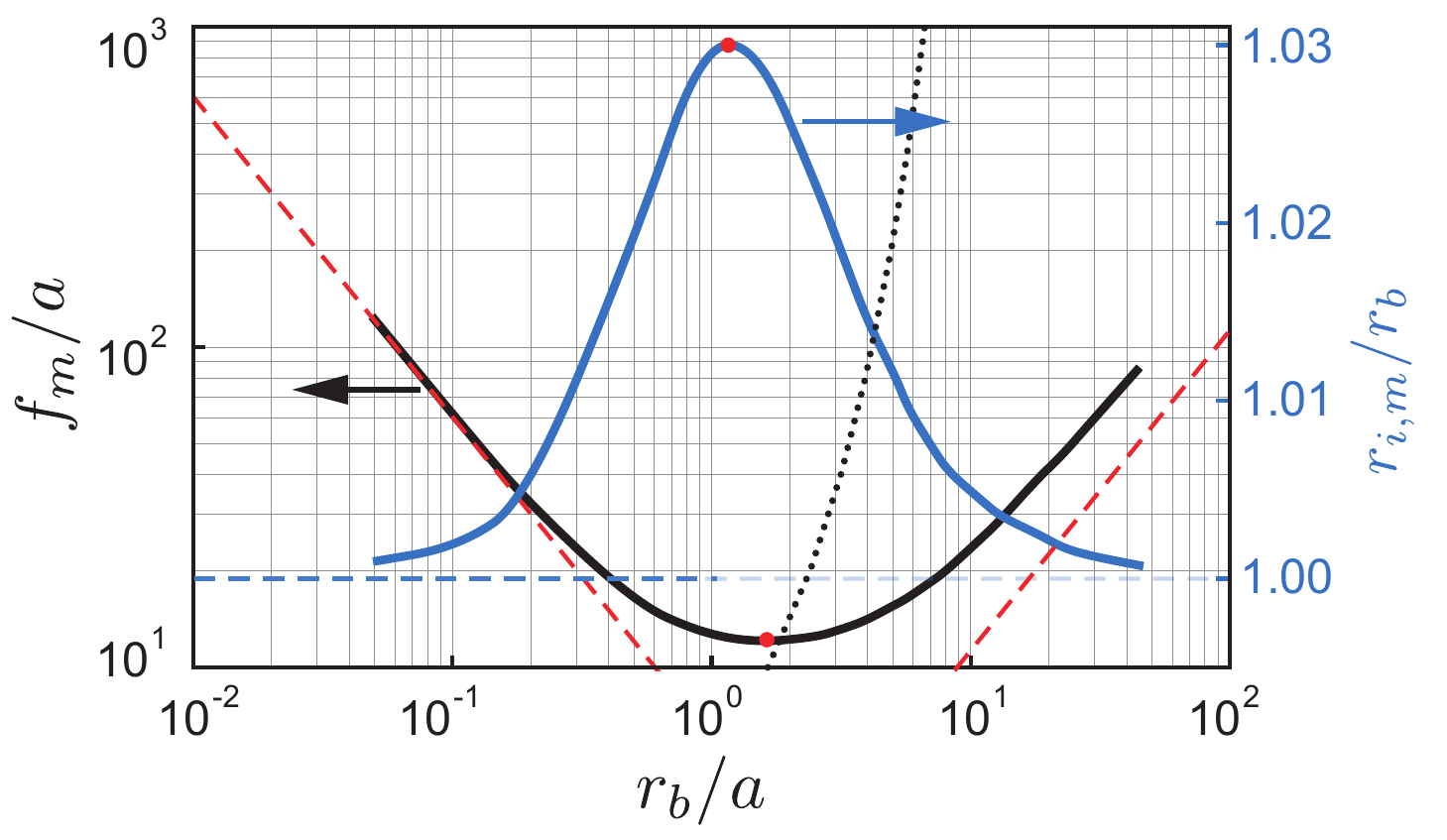}}
\caption{Relation between the bubble size $r_b/a$ and the minimal focal distance $f_m/a$ (left axis and grid lines, log-log plot). Also plotted is the impact parameter for which this minimal focal distance is realized (right axes, log-linear plot). The red dashed lines show the limiting behaviour in the small bubble regime, \eqref{eq:fapprox2}, and the large bubble regime (cf.\ text and Fig.\ \ref{Fig_SketchApprox}(b)). The blue dashed line shows the approximation $r_i/R_0\approx r_i/r_b=1$ used (blue) to derive the small bubble approximation. The dotted black line shows the paraxial focal length $|f_{bb}| \approx |R_0/(n_w^{-1}-1)|$ of the lower bubble cavity.}
\label{Fig_fmin}
\end{figure}

\begin{figure}[htbp]
\centering
\fbox{\includegraphics[width=\linewidth]{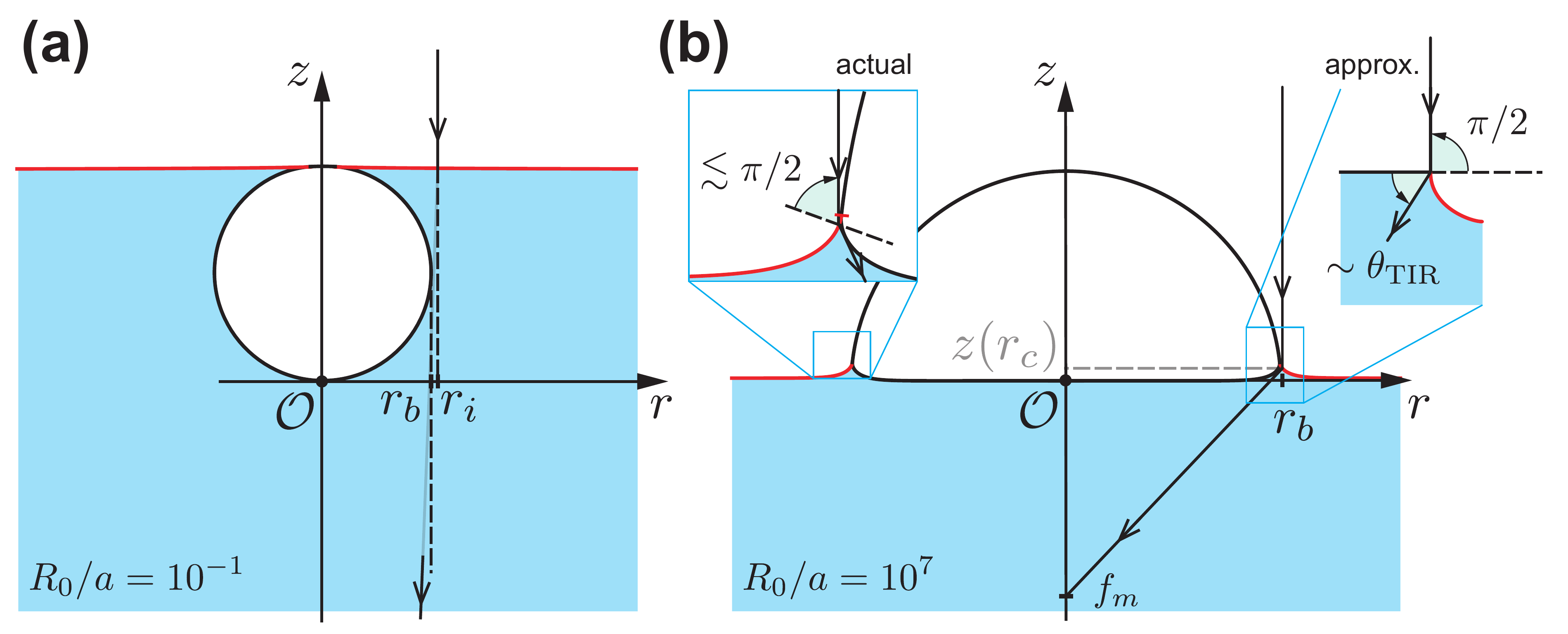}}
\caption{Sketch of the two limiting cases: \textbf{(a)} Small bubbles refract light towards the beginning of the axial caustic for $r_i\gtrsim r_{i,m}\approx r_b$. The gentle slope approximation is valid and the outer meniscus is well described by \eqref{eq:zr}. \textbf{(b)} For large bubbles, the three interfaces meet with an almost vertical slope ($z' \rightarrow \infty$ or $\phi_c\gtrsim 90^{\circ}$ for $R_0/a\rightarrow \infty$). The refracted angle towards the horizontal is close to (but larger than) $\theta_{\rm TIR}$, such that $\tan(\theta_{\rm TIR}) \lesssim f_m/r_b$.}%\lessapprox \gtrapprox
\label{Fig_SketchApprox}
\end{figure}

Now, going beyond the approximation discussed above, and considering bubbles of arbitrary sizes $r_b$, the focus starts no earlier than about $f_m = 12.1 a$ (for water: $3.3\,\rm cm$, agreeing with the observations in Fig.\ \ref{Fig_Bath}) below the water surface (at infinity) for intermediate size bubbles. This distance increases both for smaller (see previous section) but also for larger bubbles. A minimum of $f_m$ is observed around $r_b/a = 1.65$ ($R_0/a = 2.46$), i.e.\ for bubbles in water of diameters $\varnothing = 9.0 \,\rm mm$, see the black curve in Fig.\ \ref{Fig_fmin} (left axis) and the red marker. The actual impact parameter $r_i$ for which these minimum focal distances are realized are close to the bubble radii $r_b$, and at most $3.0\%$ larger (at $r_b/a = 1.15 a$ or $R_0/a= 1.36$), see the blue line and the red marker, both plotted with the corresponding right axis in Fig.\ \ref{Fig_fmin}.
In the large bubble limit when $R_0/a\gg 1$, the minimum focal distances actually scale linearly with increasing bubble radii. This may be understood from Fig.\ \ref{Fig_SketchApprox}(b), where a lower bound and crude approximation is derived which yields $f_m \gtrsim r_b \tan(\theta_{\rm TIR})$. Indeed, the red dashed line sloping upward to the right in Fig.\ \ref{Fig_fmin} (plotted against the left axis) captures the black curves' linear behaviour well and acts as a lower bound.

Under \textit{oblique incidence} the critical distance of the axial caustic formation is decreased by a geometric factor, see section \ref{sec:oblique}\ref{sec:RayTracing}.

\section{Bubble menisci as Axicons}
In the preceding section, we have seen that a bubble acts to converge light, albeit \textit{not to a single focal point}. While it is tempting to call the bubble or its outer meniscus a ''meniscus lens'', this term is, as is well known, already reserved for a special type of spherical lens: a convex-concave lens. Instead, bubbles at interfaces may actually be thought of as \textit{(negative toric) axicon lenses} \cite{McLeod1954}. Axicons are rotationally symmetric optical elements which produce images along a segment of a \textit{line on its optical axis} \cite{McLeod1954,Rayces1958, McLeod1960} instead of a point as for regular perfect lenses. Yet again, the term ''meniscus axicon'' is already taken by a certain type of double-conic convex-concave axicon lens \cite{Zhang2013}, but nonetheless one may think of the meniscus around a bubble as a \textit{bubble axicon}. The axial range in which images are formed can be found from geometrical optics by determining the interference region in which rays from opposite sides of the axicon meet. This has been done above: the interference region in axial direction extends over $[f_m,\infty)$. Due to their peculiar properties (i.e.\ the line caustic), axicons exhibit an increased focal depth \cite{McLeod1954,Rayces1958,Saikaley2013,McLeod1960,Arimoto1992,Burvall2004,Sochacki1992}. This corresponds precisely to the observations which triggered this investigation, now becoming readily comprehensible as phenomena of bubble axicons at work. The images in Fig.\ \ref{Fig_Bath} are real (inverted) images and the magnification $M=-d_i / d_o$, depends on the distance of the imaged object $d_o$ (e.g.\ light source height above the water level) and the distance of the image (i.e.\ the distance of the screen below the bubble). The magnification is hence independent of the bubble size, as is seen by the same-sized "F"-images for the differently-sized bubbles in the photo.

\section{Intensity patterns below bubbles: formation of the axial caustic}\label{sec:IntensityPatterns}
Before discussing the axial line caustic associated with this special bubble axicon lens, this section will treat the details of intensity patterns just below a bubble before the axial line caustic emerges (section \ref{sec:oblique}\ref{sec:RayTracing} will discuss the type D-ray associated phenomenology for oblique incidence using some more approximations). Applying the fundamental law of illuminance \cite{McLeod1954}, one may find the intensity in the geometrical optics limit using the density of rays and taking into account the axial symmetry:
\begin{equation}
\frac{I\left(z_p,r_p\right)}{I_0} = \frac{\Delta r_i}{\Delta r_p} \sum_{r_i\in \mathcal{S}} \overline{|t_i|^2}\, n_w \frac{\lvert r_i^2- (r_i+\Delta r_i)^2\rvert}{\lvert r_p\left(r_i\right)^2 - r_p\left(r_i+\Delta r_i\right)^2\rvert}\label{eq:intApprox},
\end{equation}
where $\mathcal{S}$ is the set of all discretely sampled impact parameters $r_i=i \times \Delta r_i$, with $i\in 1,\dots N_i$ such that the exit ray in the plane of the screen at $z_p$ has a radial coordinate of absolute value $|r_p\left(z_p,r_i\right)|$ in the range $[r_p, r_p+\Delta r_p]$, where $\Delta r_p$ is the resolution of the radial pattern hereby computed and chosen such that $\Delta r_i\ll \Delta r_p$. The significance of the absolute value, i.e.\ the allowance of $r_p=\pm r_p$ accounts for the fact that rays may come from opposite sides of the axis to reach a given point (cf.\ Fig.\ \ref{Fig_Raytracing}). The factor $\overline{|t_i|^2}$ in \eqref{eq:intApprox} is the product of all involved (Fresnel) field amplitude transmission coefficients (polarization-averaged), taking into account the relevant incidence angles to the interface normals encountered by a given ray. The factor $n_w$ is the ratio of impedances of water and air and enters the expression due to $I\propto n |E_0|^2$ (the cosine-factor involved in the (Fresnel) power transmission coefficients $T$ is already taken account for by the geometric factor of changing annulus areas).

\begin{figure*}[htbp]
\centering
\fbox{\includegraphics[width=\linewidth]{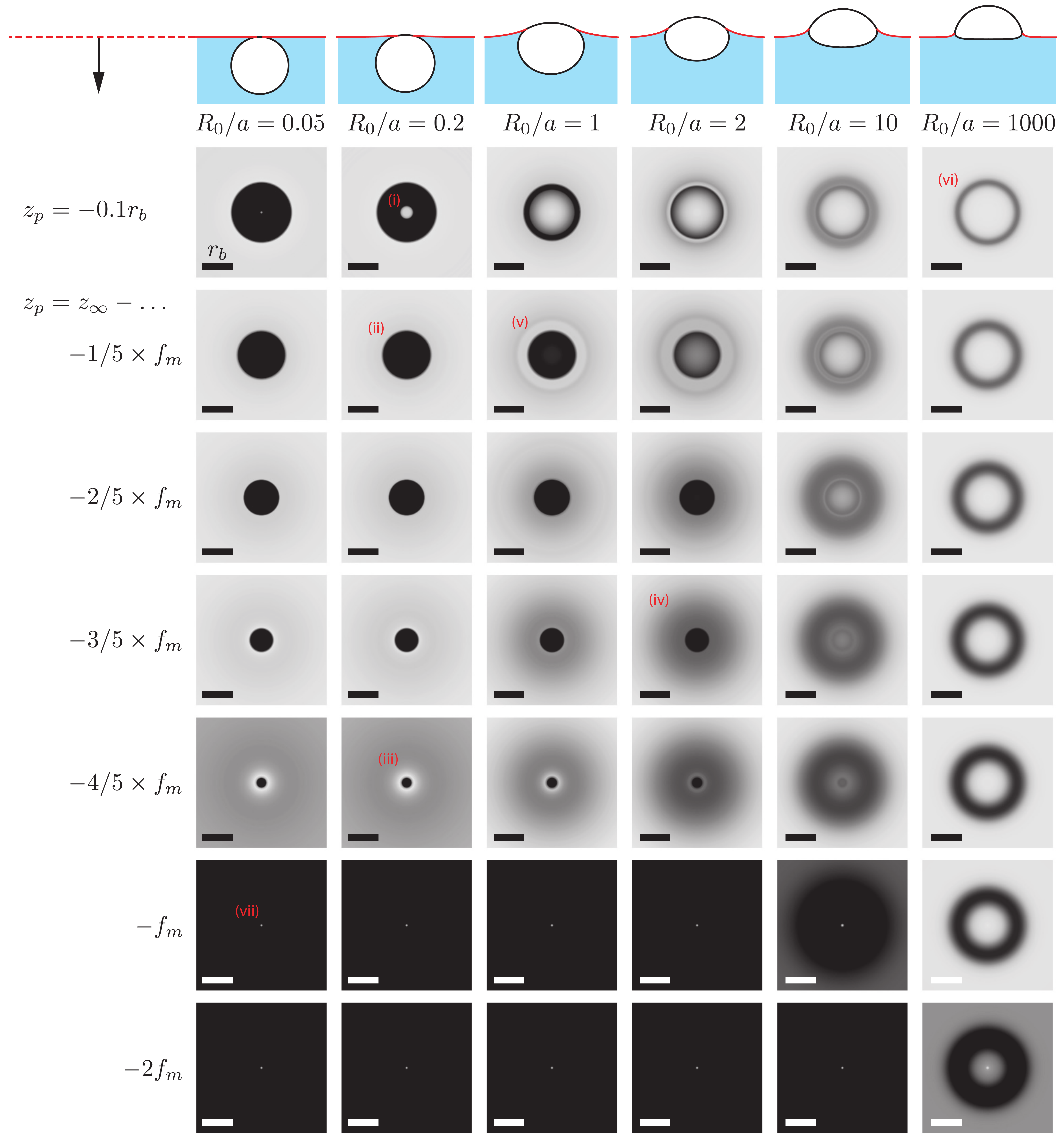}}
\caption{Intensities in planes (cf.\ also the markings in Fig.\ \ref{Fig_Raytracing}) $z_p=-0.1 r_b$, $z_\infty - 1/5\times f_m$, $z_\infty - 2/5\times f_m$, $z_\infty - 3/5\times f_m$, $z_\infty - 4/5\times f_m$, $z_\infty - f_m$, $z_\infty - 2\times f_m$ obtained via \eqref{eq:intApprox} and using $r_i\in [0, 5r_b]$, $N_i=20.000$, $\Delta r_i=5r_b/N_i$, $N_p=N_i/400$, $\Delta r_p=3 r_b/N_p$. The color scales of the images (sides lengths: $6r_b/\sqrt{2}$) are adjusted per image from $0$ (black) to $max(I)$ (bright). Several characteristic features have been labeled with lowercase roman numerals (i-vii) in red. A discussion of those and more details are given in the text.}%Fig_Geom
\label{Fig_Intensities}
\end{figure*}

Fig.\ \ref{Fig_Intensities} shows a series of intensity plots at various depths below the water surface, starting just below the bubble ($z_p=-0.1r_b$) and then proceeding in fractions of the minimal focal distance until reaching $z_p=z_\infty - f_m$, and finally the intensity in a plane at $z_p=z_\infty - 2 \times f_m$. The axial caustic clearly shows up after the minimal focal distance, and exists in all planes below this depth. This general observation agrees with the work of Adler et al.\ \cite{Lock2015} for the case of floating symmetric objects under an external pull, cf.\  loc.\ Fig.\ 5 and 13.

The intensity distribution just below the bubble can be seen to be composed of a \textit{central bright disc of radius $r_c$} (light is transmitted without refraction through the cap), and a \textit{dark annulus around it} (light is diverted away in either direction due to the interior and outer menisci) (\textit{feature (i)}). This dark ring of almost unity contrast disappears for larger bubbles somewhere around $R_0/a\sim 1\dots 2$: In the ray tracings, Fig.\ \ref{Fig_Raytracing}, this dark annulus can be identified with the region of sparse ray density outside of the cap perimeter $r_c$ just below the bubble before the inward refracted type D rays overlay with the outward refracted type A rays (for $r_i=r_c$).

For small bubbles with $R_0/a\ll 1$, i.e.\ the first two columns, a \textit{high-contrast dark shadow} region emerges not far from the bubble, and the central bright disc disappears quickly (\textit{feature (ii)}). As compared to larger bubbles, the shadow is relatively sharp and appears against a fairly homogeneous intensity background. The reason for this is that the small-bubble meniscus which originates from the cap at $r_c \ll R_0$ has decayed already appreciably before the type D rays that just miss the bubble with $r_i\approx r_b\approx R_0$, and which form the edge of the shadow, are much less refracted than those shadow rays for larger bubbles. The central spot's quick disappearance is due to the short focal length $f_{bb}\approx R_0/(n_w^{-1}-1)$ of the bubble's bottom (and measured relative to it) acting as a diverging lens. Nonetheless, the shadow shrinks in size, since the outer meniscus acts to weakly converge the rays that miss the bubble, as shown in the ray tracings of Fig.\ \ref{Fig_Raytracing}. This is the shadow shrinking already described for small floating objects by Berry and Hajnal in \cite{Berry1983} (cf.\ also Ref.\ [\!\!\citenum{Walker1988}]). The authors noted (in concrete reference to a floating edge, albeit they discuss spheres in parallel) that in this scenario no real caustics are involved: This statement is now seen to hold only in the domain above $f_m$, whereafter the line caustic emerges. Another feature appears when the inward refracted type D ray's radial convergence overcomes the angular divergence of those same rays: A \textit{bright halo around the shadow} (\textit{feature (iii)}) signifies the increased ray density here caused by their convergence.

For larger spheres, i.e.\ columns three to five ($R_0/a = 1$ to $R_0/a=10$), a \textit{dark but smooth halo} around the direct shadow becomes prominent (\textit{feature (iv)}). It signifies the diverting action of the bubble's bottom (now acting on a significant portion of $r_i<r_b$) as well as the ray-diverting (and divergence-imprinting) action of the outer meniscus (now strongly affecting type D rays). The missing irradiance in this dark halo is thus either diverted and redistributed to the periphery or contributes to the lit annulus inside the former shadow (i.e.\ shrinking the shadow). In planes closer to the bubble, the superposition of type A and D rays can also create a \textit{bright ring} (\textit{feature (v)}).

For very large bubbles $R_0/a\gg 1$, i.e.\ those which resemble a half-dome with only small (relative to the bubbles size) menisci inside and outside the bubble rim at $r_c\approx r_b$, a \textit{dark and smooth ring shadow} (\textit{feature (vi)}) of the menisci appears below the rim, having non-unit contrast. It is an extreme case of the dark halo discussed above, but with a central discernible bright disc region due to the flatness of large bubble bottoms. The enclosed bright area is a fuzzy version of the bright and sharp unit-contrast cap-transmission disk discussed for small bubbles: here, however, the cap shadow rays coincide with the inner meniscus shadow rays and thus get refracted significantly to form a smooth shadow edge. The dark ring broadens with increasing depth, as shown in the right-most column. This dark ring has a non-unit contrast which grows with depth (not shown in the images, since they are each min-max color-scaled). Again, this behaviour is easily reconciled with the ray tracing shown in the right-most scenario of Fig.\ \ref{Fig_Raytracing}.

Finally, the axial caustic appears (\textit{feature (vii)}) at and below a depth $f_m$. The contrast at its first appearance decreases for increasing bubble sizes, as can be seen in the last two rows of Fig.\ \ref{Fig_Intensities} from left to right each.

The above analysis and the intensity patterns for screen planes above $f_m$ likely explains why some blurred images could already be observed at depths of only $\sim 2.5 - 3 \rm cm$ (instead of only below $3.3\,\rm cm$, cf.\ section \ref{sec:minfocdistance}) in the experiment reported in Fig.\ \ref{Fig_Bath}. The bright halo (feature (iii)) around the shrunken shadow acts as an annulus focus to form blurred images that were only just discernible.

\section{Axial intensity} \label{sec:AxialInt}
\subsection{Geometrical Optics}
Previous studies on axicon design in the limit of geometrical optics assumed $P_z\left(z\right) \mathrm{d}z = 2\pi P_\sigma(r) r_i \mathrm{d}r_i$ to hold \cite{Sochacki1992,Wang2017}. The interpretation of this expression being that "the two-dimensional power density $P_\sigma(r)$ (in units of power [per area]) is being transformed (squeezed) into the one-dimensional axis density $P_z(z)$ (in units of power [per length])", and that the "quantity $P_z(z)$ in this formulation [could] be interpreted as a first approximation of the in-axis light intensity $I(r=0,z)$ that would result from a diffraction integral." \cite{Sochacki1992}. From this expression, and considering that the rays carrying the power make an angle $\theta_f$ to the axis \cite{Wang2017}, one finds (now writing $z=z_p$)

\begin{equation}
\frac{I_z\left(z\right)}{I_0} = \overline{|t_i|^2}\, n_w \frac{2\pi r_i(z)}{\mathrm{d}f_i(r_i)/\mathrm{d}r_i|_{r_i(z)}} \cos\left(\theta_f(r_i(z))\right)\label{eqn:I}
\end{equation}

\begin{figure}[htbp]
\centering
\fbox{\includegraphics[width=0.7\linewidth]{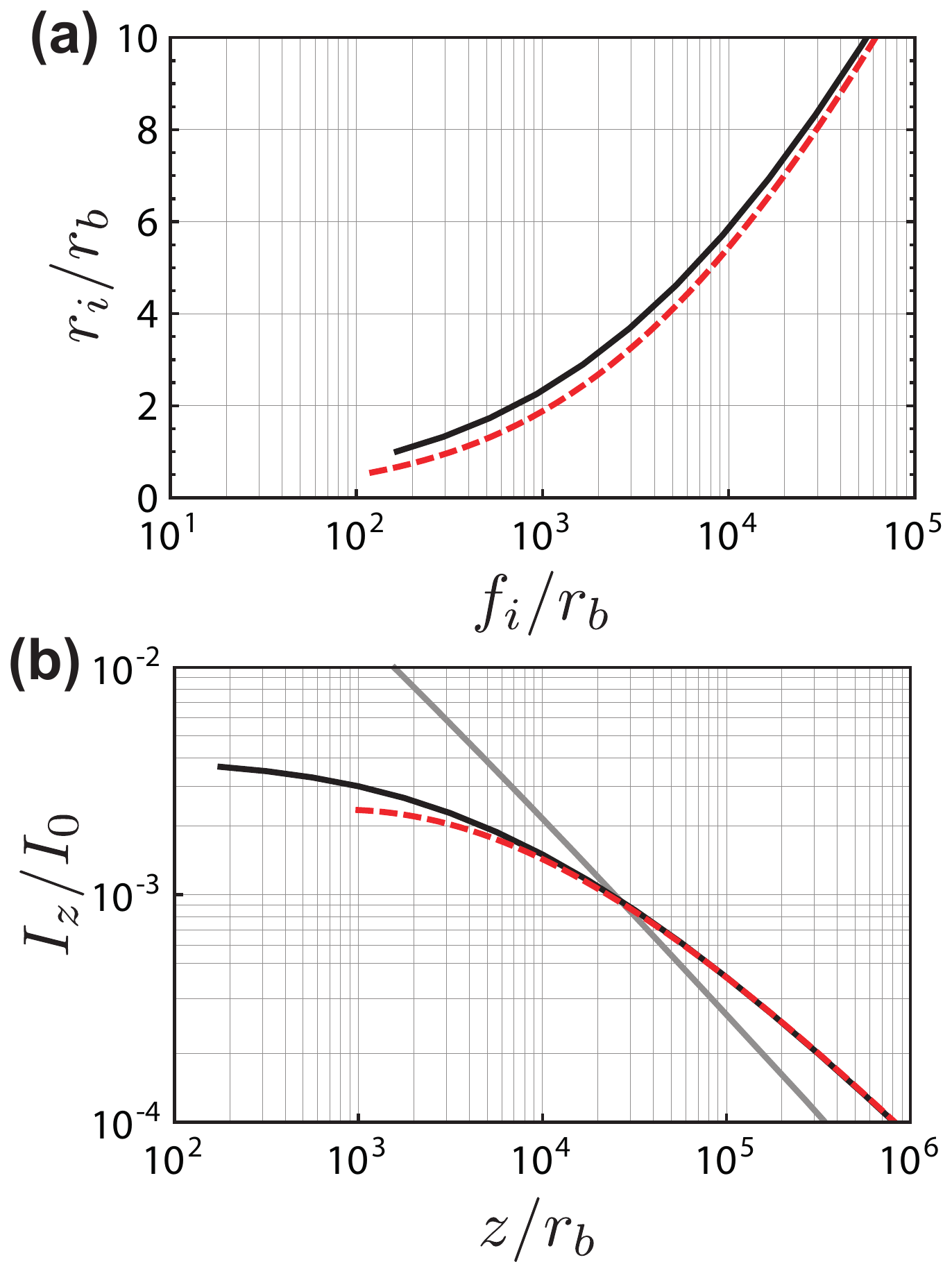}}
\caption{\textbf{(a)} Inverse of the relationship: axis intercept for refracted rays vs.\ their impact parameter $r_i$. The red dashed lines shows the first order approximation from \eqref{eq:fapprox}. \textbf{(b)} Corresponding axial intensity from \eqref{eqn:I} and the approximation \eqref{approx:I} (red dashed line) for $R_0/a=0.2$. The grey line shows a larger bubble $R_0/a=2$.}%\lessapprox \gtrapprox $I_z/(I_0 n_w)$
\label{Fig_Axicon}
\end{figure}

Considering again the analytically tractable small-bubble limit, and inverting \eqref{eq:fapprox} in the large $r_i$-limit for $r_i(z)$, one finds approximately an axial intensity decay described by:
\begin{equation}
\frac{I_z\left(\bar{z}\right)}{I_0 } \sim \frac{3\pi W(J)^2}{\bar{z}+\bar{z} W(J)},\quad J=\frac{4}{3}\pi^{\frac{1}{3}}\left(\frac{R_0}{a}\right)^2\left[\frac{(n_w-1) \bar{z}}{n_w(6-R_0^2/a^2)}\right]^{\frac{2}{3}}\label{approx:I}
\end{equation}
where $\bar{z}=z/a$, $W(x)$ is the Lambert W-Function (ProductLog-function), i.e.\ the inverse of $x\exp(x)$. The cosine factor and the amplitude transmission factors (including the impedance ratio $n_w$) were ignored since they are roughly constant and of order unity. The expression shows a slow decay of the axial intensity in the ideal scenario described by \eqref{approx:I}. Although the approximation function \eqref{approx:I} has a local maximum, the exact expression \eqref{eqn:I} using the values from raytracing shows a monotonous decay only. On an absolute length scale, larger bubbles appear to show a steeper decay, although the general observation remains: the intensity remains significant even after axial distances of many bubble radii. This is qualitatively consistent with observations (cf.\ also Fig.\ \ref{Fig_Pool}), although the divergence of most light sources and non-vertical light incidence will deteriorate the axial focus quality as described in section \ref{sec:oblique}\ref{sec:CatastropheTheory}.

\subsection{Diffraction}
The axial intensity pattern could also be found in wave optics. In the work of Berry and Hajnal \cite{Berry1983}, the corresponding diffraction integral was analyzed mostly for depressed water surface deformations. Note that their loc.\ eq.\ (30) for the gentle slope approximation around a floating sphere corresponds to \eqref{eq:zr} with $D\rightarrow a C < 0$ (i.e.\ $H<0$). Their loc.\ eq.\ (38) should still capture the caustic below a bubble well in the scalar diffraction approximation, neglecting the action of the diverging bubble cavity. The contribution due to the \textit{increasingly flat} central bubble cavity transmission for $r_i<r_c$ is expected to be insignificant only beyond $f_{bb}$, where $|f_{bb}| > f_m$ (for $R_0/a \gtrsim 3$, cf.\ Fig.\ \ref{Fig_fmin}), as can be seen by the low contrast (in the respective plots towards the lower right) in Fig.\ \ref{Fig_Intensities} of the central caustic against the backdrop of the bright halo / direct transmission. This is different from the transparent round (rigid) sphere case discussed in \cite{Berry1983}, where it was remarked that significant contributions are expected only close to the sphere's ($n>1$) focus ($z_p \sim - R$, cf.\ loc. p.\ 34). A detailed quantitative discussion of the diffraction integral of the caustic is beyond the scope of this paper, albeit the next section gives some general qualitative remarks.

\section{Oblique incidence}\label{sec:oblique}%spun cusp: p. 196 of nye book 
Up to this point only vertical illumination was considered: the angle $\Gamma$ of the incidence light to the vertical was assumed to be zero, especially in sections \ref{sec:IntensityPatterns}-\ref{sec:AxialInt}. The following subsections focus on the phenomenology and implications for the caustic of oblique incidence with $\Gamma >0^{\circ}$. For sun light, the incident angle is related to the solar elevation $e$ via $\Gamma = 90^{\circ}-e$.
\subsection{Catastrophe Theory: Unfolding of the line caustic}\label{sec:CatastropheTheory} %spun cusp: p. 196 of nye book
For a regular convergent lens (when illuminated normally on-axis) with spherical aberration there are two (real) caustics close to the paraxial focus: the so-called spun cusp (a rotationally symmetric cusped cone) and a line segment caustic, both meeting at the primary focus \cite{Berry1980,Berry1981,Nye2005}. For the common refracting cone axicon, only the axial caustic remains real whereas the second spun cusp caustic becomes a mere virtual surface \cite{Rayces1958}. The same holds true for the bubble axicon, which also only forms a ''naked'' axial caustic. Rayces, in 1958, already qualitatively described how the image of an \textit{off-axis} point source for an axicon unfolds into a four-cusped caustic (cf.\ Fig.\ 4 in Ref.\ [\!\!\citenum{Rayces1958}], also Ref.\ [\!\!\citenum{Arimoto1992}]), which McLeod later constructed geometrically to show its \textit{astroid} shape (4-cusped hypercycloid / scaled version of the parametric curve $(\sin^3 t, \cos^3 t)$, referred to as a "kite") \cite{McLeod1960}. For oblique incidence or perturbed bubbles the same fate will unfold the axial caustic described in the earlier sections.

\begin{figure*}[htbp]
\centering
\fbox{\includegraphics[width=\linewidth]{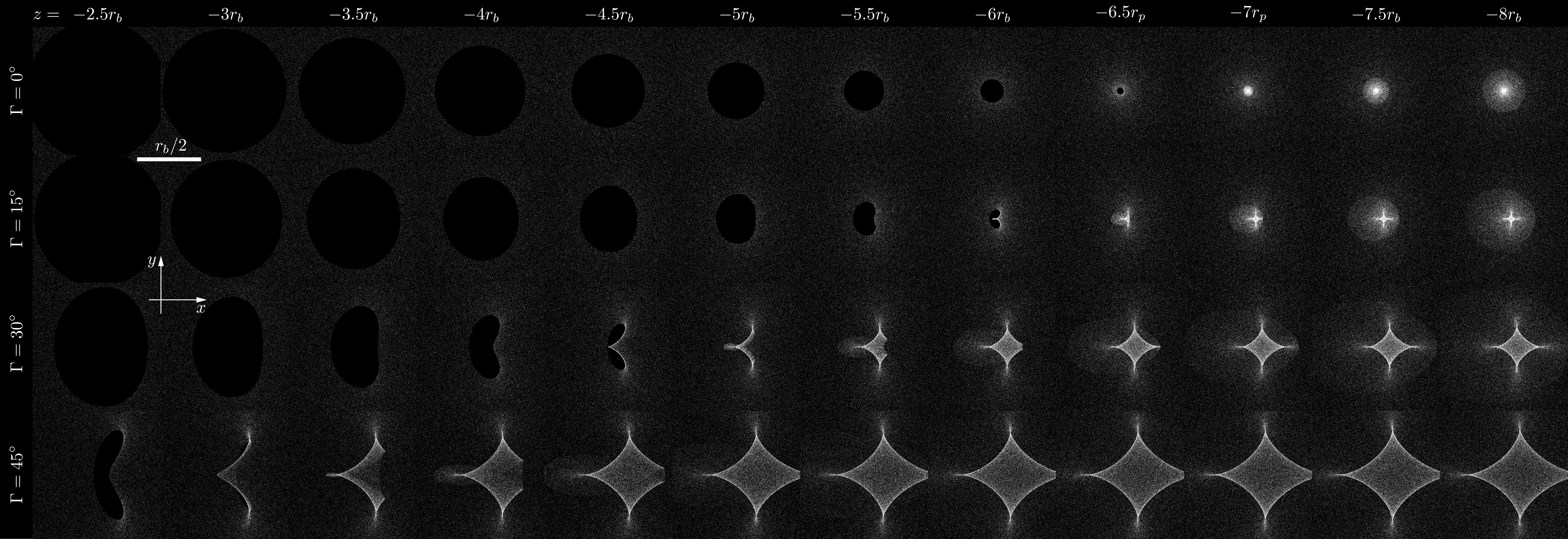}}
\caption{Ray tracings ($1\rm M$ rays with $(x_i,y_i)\in [-2.4r_b,2.4r_b]$) for $R_0/a=2$ showing the emergence of the bubble axicon caustic with increasing distance (left to right) of the projection screen to the bubble (gray depth marks for $R_0/a=2$ in Fig.\ \ref{Fig_Raytracing}, from $z/(f_m-z_\infty)=-0.3$ to $-1.1$). The ray tracings have been done for three incident light directions $\Gamma$ (top to bottom rows), showing plots of side lengths $r_b \times r_b$ (scale as inset). For increasing $\Gamma$, the fully formed caustic appears closer to the bubble, see also Fig.\ \ref{Fig_SketchEmergence}(a).}%\lessapprox \gtrapprox %and assume perfectly parallel incoming light (no divergence)
\label{Fig_Formation}
\end{figure*}

In the context of catastrophe theory, which was not available at the time of those early studies, this can in the meanwhile understood as part of a natural unfolding of the \textit{two unstable caustics} related to the primary (Seidel) aberrations \cite{Berry1980,Berry1981}: Both the line caustic and the spun cusp are non-elementary (non-generic) and thus structurally unstable caustics, meaning that they do not belong to the class of generic and classifiable caustics which are topologically stable against perturbations. In particular, they both are, in the language of catastrophe theory, caustics of infinite codimension, meaning that an infinite number of perturbation parameters (say the Fourier coefficients in an expansion) are required to prescribe an unfolding perturbation \cite{Berry1976}, and that there are consequentially an infinite number of ways it can in principle unfold. In the case where both the spun cusp and the axial line caustic unfold simultaneously, and returning to the case of primary aberrations only (imprinting special types of perturbations on the wavefront), a structurally stable (containing elementary catastrophes only) topology emerges in which two hyperbolic umbilic foci (/singularities) are embedded in a global caustic structure that contains the four-cusp (and four folds) figure and an almost conical fold-surface. Berry and Upstill (appendix 2 of Ref.\ [\!\!\citenum{Berry1980}], based on notes of Hannay) show a sequence of sections through this topology, and Berry's lecture notes \cite{Berry1981} give a three-dimensional depiction in Fig.\ 41. Nye shows its embedding in the unfolding of the higher-order catastrophe $X_9$ for two-fold symmetry and modulus $+2$ (Fig.\ 8.3 in Ref.\ [\!\!\citenum{Nye1999Book}], Fig.\ 21 of Ref.\ [\!\!\citenum{Nye1986}], cf.\ also Appendix B in Ref.\ [\!\!\citenum{Nye2018}]). This structure and the associated hyperbolic umbilic foci are also encountered in spheroidal drop \textit{rainbow scattering} \cite{Marston1984,Nye1984} or \textit{thin drops} in rectangular or rhombus-shaped apertures \cite{Nye1986}. When the spun-cusp is missing (as for our bubble axicon, and other cases, see below), only the central fold surfaces with needle- or star-like cross-sections emerge (the astroid), and no hyperbolic umbilic foci exist.

The astroid caustic also occurs isolated in the context of \textit{glory scattering} by a spheroidal particle caused by the correspondingly perturbed toroidal wavefront \cite{Marston1989} (cf.\ loc. Fig.\ 1(b) and 7(a), here as a 4 to 2 ray transition with 1 extra background ray on top of the 3 to 1 ray transition characteristic of a cusp-fold). %It may further be observed for square-like (or rhomboidal) perturbations of thin (sessile or pendant) water drop lenses \cite{Berry1976,Nye2018} (nye?)

A most impressive natural demonstration of the naked axial caustic unfolding has been observed in 1976 when the star $\epsilon$ Geminorum was occulted by Mars \cite{Elliot1977,Berry1981}: the \textit{atmosphere of Mars acted as a toroidal axicon lens} (an annular lens). If Mars had been perfectly spherical, an axial line focus would have formed. Instead, Mars (like any spinning planet) and its atmosphere is well described by an oblate spheroid, hence perturbing and unfolding at the time the line caustic in one of the simplest ways geometrically possible into the four-cusped astroid caustic of roughly $150\,\rm km$ diameter as projected onto earth. Similar observations have later been made for instance during an occultation of the star 28 Sgr by Saturn \cite{Nicholson1995}, or at radio frequencies for the spacecrafts Voyager 1 and 2 passing behind by Jupiter \cite{Eshleman1979}. Four images exist for an observer passing the interior of the astroid caustic, while two exist just outside of it. The same holds true for solar gravitational lensing \cite{Loutsenko2018}. Much more general, gravitational lensing by aspherical (e.g.\ elliptic) astronomical lenses shares many commonalities (including the central astroid caustic) with atmospheric lensing \cite{Nye1999Book,Petters2001book}, although here one has a transition from 5 to 3 to 1 images, cf.\ chapter 3.2 and Fig.\ 3.16 of Ref.\ [\!\!\citenum{Petters2001book}] (the unfolded line caustic is not naked and embedded in the exterior elliptical fold caustic). Hence, the closest astronomical analogy probably exists to the atmospheric lensing situation: Similar to the ellipsoidally perturbed annulus lens of a planet's atmosphere, the bubble axicon lens when viewed at an angle deviates from the circular symmetry and appears ellipsoidal. Accordingly, the astroid unfolding of the naked axial caustic is expected, and similarly a 4 to 2 images transition when viewing light sources through it. This, of course does not mean that the individual cusps involve more than their characteristic coalescence of 3 rays, and inherently present background rays make these higher-order caustics awkward to analyze. %A ''terrascope'' utilizing the atmospheric lensing of the Earth has recently been suggested, too \cite{Kipping2019}, and the same phenomenology would occur.

As a side note, as is well known, an axicon lens similar to the stem of a wine glass may be used to simulate the effects of gravitational lensing \cite{Surdej1993,Lohre1996}. Unfortunately, the macroscopic bubble axicon's shape (exponential in its large-distance fall-off) is no better approximation to the logarithmic profile needed to generate a proper optical analogon to a point mass gravitational lens (i.e.\ to gravitational microlensing), nor presumably to other trivial mass distribution lenses \cite{Surdej1993}. The microscopic limit, however, indeed allows a perfect analogy to the point mass as here the meniscus profile becomes $\Delta z(r) \propto -\ln(r/2a)$ \cite{DiLeonardo2003,Hennequin2013}.

Based on the scaling laws of diffraction catastrophe theory \cite{Berry1980,Berry1981}, and here inferring from the singularity index associated with the corresponding elementary catastrophes, the light intensity close to the cusps (where two folds meet tangentially) should rise to $\mathcal{O}(k^{1/2})$, whereas it should rise to $\mathcal{O}(k^{1/3})$ approaching the folds. Similar predictions on the diffraction fringe spacings in the astroid may be inferred from the generic theory \cite{Berry1980,Berry1981}. 

In summary, the bubble axicon under oblique illumination is another prime natural example for the simplest possible unfolding of the unstable line caustic.% and consequentially shares many aspects of its phenomenology with other lenses occuring in nature.

\subsection{Monte Carlo vector ray tracing and the astroid's evolution}\label{sec:RayTracing}
\begin{figure}[htbp]
\centering
\fbox{\includegraphics[width=1\linewidth]{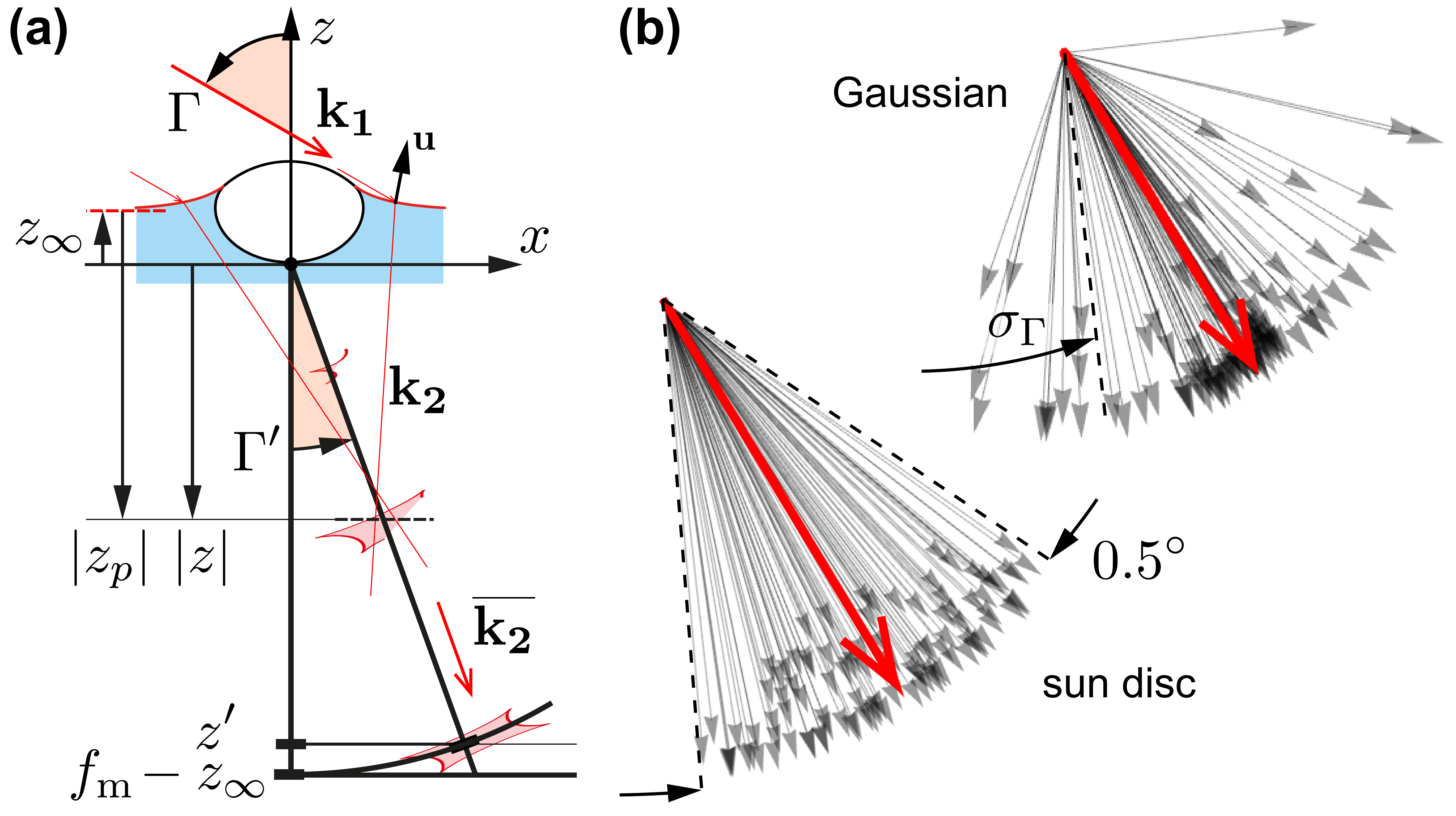}}
\caption{\textbf{(a)} Sketch motivating the geometric factor $\cos(\Gamma')=z'/(f_m-z_\infty)$ describing the decreased depth $z'$ at which the unfolded focus appears under oblique illumination. Here, $\sin(\Gamma)=n_w \sin(\Gamma')$ and $\overline{\mathbf{k_2}}={(\sin(\Gamma'),0,-\cos(\Gamma'))}$. Moving the projection plane downwards in $z$, the coalescing rays from the $+x$-side of the bubble form the $-x$-cusp first (cf.\ section \ref{sec:sausage} for details on the inversion). \textbf{(b)} Illustration of the two scenarios for the divergence of light considered in section \ref{sec:divergence} and Fig.\ \ref{Fig_Fade}. The divergence was each scaled by $100\times$ for this sketch.}%\lessapprox \gtrapprox Sketch of the rotations applied to the incidence ray vector $\mathbf{k_1}$ (read from top left to bottom right) to simulate light divergence.
\label{Fig_SketchEmergence}
\end{figure}

To further study the bubble axicon's astroid caustic, Monte Carlo vector ray tracing simulations were done similar to those in \cite{Lock2015}. This, for instance, allows some characteristics of the emergence and shape of the axial caustic to be described. The incident ray vectors were set to $\mathbf{k_1}=(\sin(\Gamma),0,-\cos(\Gamma))$ (cf.\ Fig.\ \ref{Fig_SketchEmergence}(a)), and the refracted ray vectors $\mathbf{k_2}$ then computed via loc.\ eq.\ (42) of Ref.\ [\!\!\citenum{Lock2015}] (vector form of Snell's law of refraction). The outward-pointing unit surface normal vectors $\mathbf{u}=\mathbf{u'}/|\mathbf{u'}|$ at points $\mathbf{m_i}=(x_i,y_i,f)$ were calculated using finite differences and $\mathbf{u'}=(\partial_x f,\partial_y f,-1)$, with the outer meniscus surface profile $f(x_i,y_i)=\Delta z(r_i)$ taken with cartesian coordinates related to impact parameters through $r_i^2=x_i^2+y_i^2$. The intersections $\mathbf{r}$ of lines $\mathbf{r}=\mathbf{m_i} + \lambda \mathbf{k_2}$ with the projection plane $[\mathbf{r}-(0,0,z)]\cdot \mathbf{\hat{z}}=0$ are then computed via $\mathbf{r}=\mathbf{m_i}+\mathbf{k_2}([(0,0,z)-\mathbf{m_i}] \cdot \mathbf{\hat{z}})/(\mathbf{k_2}\cdot\mathbf{\hat{z}})$ and visualized in scatter plots. Since the meniscus acts likewise as an axicon for all bubble shapes, the same phenomenology is expected for any $R_0/a$. In the following discussions and plots, $R_0/a=2$ was chosen, corresponding to a $\varnothing 8\,\rm mm$ air-bubble on water. For simplicity, and in light of the complications associated with a correct treatment for oblique incidence, the impact parameters were sampled randomly with $r_i\ge r_b$, and were all assumed to be of type D, thereby neglecting rays of type A-C. By the foregone analyses, no qualitative (though slight quantitative) discrepancies are expected due to this simplification (cf.\ the blue curve in Fig.\ \ref{Fig_fmin}).

Fig.\ \ref{Fig_Formation} now shows the result of such ray tracing simulations. Similar to the corresponding column in Fig.\ \ref{Fig_Intensities} for $R_0/a=2$, the upper sequence depicts the emergence of the central line caustic. As noted above, due to the consideration of all $r_i>r_b$, the exact location of the starting point of the focus is slightly off and appears already for $z \sim -6.7 r_b$, whereas the actual focus distance is $f_m-z_\infty=7.6 r_b$ (cf.\ Figs.\ \ref{Fig_fmin} and \ref{Fig_Intensities}). Nonetheless, the phenomenology is clear and the shadow shrinking well visible. Also the bright halo is well-discernible, whereas the dark halo structure is too large for it to be noticeable in these plots of side lengths $r_b\times r_b$ (cf.\ Fig.\ \ref{Fig_Intensities}, showing the intensity patterns in plots of side lengths $\sim 4.2r_b \times 4.2r_b$).

\begin{figure*}[htbp]
\centering
\fbox{\includegraphics[width=\linewidth]{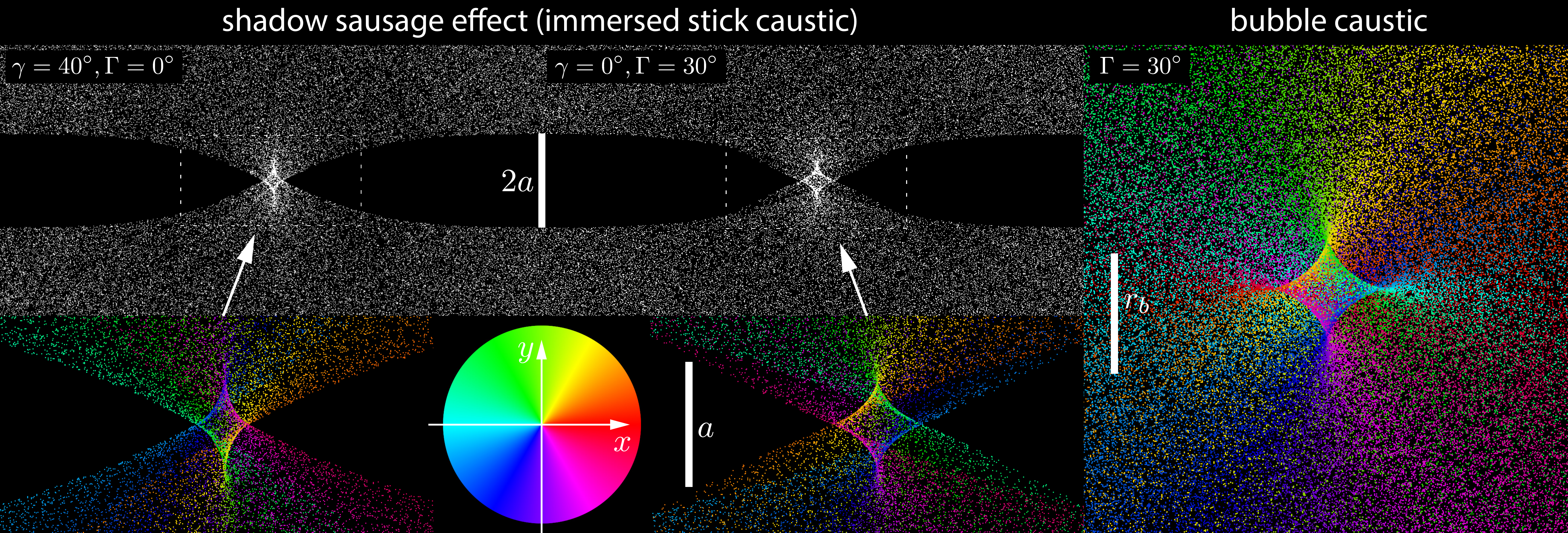}}
\caption{Comparison of the caustic structures of the shadow sausage effect (left) and the bubble caustic from ray tracing. For the inclined rod, menisci as described analytically in Ref.\ [\!\!\citenum{Lock2003}] were used, with $q=0.1$ quantifying the capillary rise difference in the obtuse and acute regions. The dots showing individual ray endpoints intersecting the projection plane (at $z=-100 r_b$, $R_0/a=2$) have been color-coded by hue according to their azimuth origin on the meniscus around the rod or bubble (see color wheel). The angles $\gamma$ and $\Gamma$ are the angles of the rod relative and the incidence light relative to the vertical, respectively. For the tilted rod (left-most plots) the $y$-cusps are inverted, while for both the bubble caustic and the vertical rod the $x$-cusps are inverted (although for the rod, the $x$-cusps lie in the shadow). Experiments with the setup shown in Fig.\ \ref{Fig_Oblique}(a) confirmed these predictions.}%\lessapprox \gtrapprox
\label{Fig_FalseColor}
\end{figure*}

The other sequences for $\Gamma=15^{\circ}, 30^{\circ}$ and $\Gamma=45^{\circ}$ (second to last row) show how the axial caustic, unfolded into the astroid caustic, emerges from a shrunken and perturbed shadow to finally morph into the four-cusped shape. A similar behaviour (minus the astroid) was reported for the tilted flat model leaf system in \cite{Lock2015}, a scenario with a half-raised and half-depressed water surface perturbation ($\Gamma=0^{\circ}$, loc. Fig.\ 7(a)-(c)). Since here the projection screen intersects the unfolded axial caustic (i.e.\ the fold surfaces forming the astroid) at an angle not perpendicular to its symmetry axis, as sketched in Fig.\ \ref{Fig_SketchEmergence}(a), the caustic appears closer in $z$ towards the bubble relative to the $\Gamma=0^{\circ}$-scenario approximately by a factor of $1/\cos(\Gamma')$, where $\Gamma'$ is the average refraction angle related to the incidence direction via $n_w \sin(\Gamma')=\sin(\Gamma)$. This average refraction angle was also used to set the center for the ray tracing plots (intersection of a hypothetical ray refracted by a level water surface at $z=z_\infty$, although the astroid is offset by a small amount in the $+x$-direction from the hereby determined center \cite{Lock2003}). For the same geometrical reason, the caustic features corresponding to rays emanating from the $+x$-side of the bubble's meniscus appear first, thus causing the \textit{$-x$-astroid cusp to appear first} (cf.\ the $x$-inversion discussed in section \ref{sec:sausage} and shown in Fig.\ \ref{Fig_FalseColor}). Correspondingly dedicated observations in a white ceramic bowl and soap-water on a sunny day confirmed the general phenomenology.

\subsection{Effect of the divergence of the illuminating light}\label{sec:divergence}
So far, perfectly parallel light (point light source at infinity) has been assumed for illumination. However, for instance the divergence of the sun's rays of about $\Delta \Gamma \sim 0.25^{\circ}$ (half angle) causes the caustic pattern to become \textit{diffuse and blurred after some distance}. 

Divergence of light was added to the vector ray tracing simulations (cf.\ section \ref{sec:RayTracing}) by successive rotation operations acting on each vector $\mathbf{k_1}$ to yield randomly perturbed ray incidence vectors $\mathbf{k_1}'$, i.e.\ $\mathbf{k_1}'=R_{\mathbf{\hat{y}},-\Gamma}\cdot R_{\mathbf{\hat{z}},{\rm rand}[0,2\pi]}\cdot R_{\mathbf{\hat{x}},\phi_s}\cdot R_{\mathbf{\hat{y}},\Gamma}\cdot \mathbf{k_1}$ with $R_{\mathbf{v},\alpha}$ being 3D rotation matrices for a counterclockwise rotation around the vector $\mathbf{v}$ by an angle $\alpha$ and $\left\{\mathbf{\hat{x}},\mathbf{\hat{y}},\mathbf{\hat{z}}\right\}$ being the axes' unit vectors, see Fig.\ \ref{Fig_SketchEmergence}(b).

\begin{figure}[bthp]
\centering
\fbox{\includegraphics[width=\linewidth]{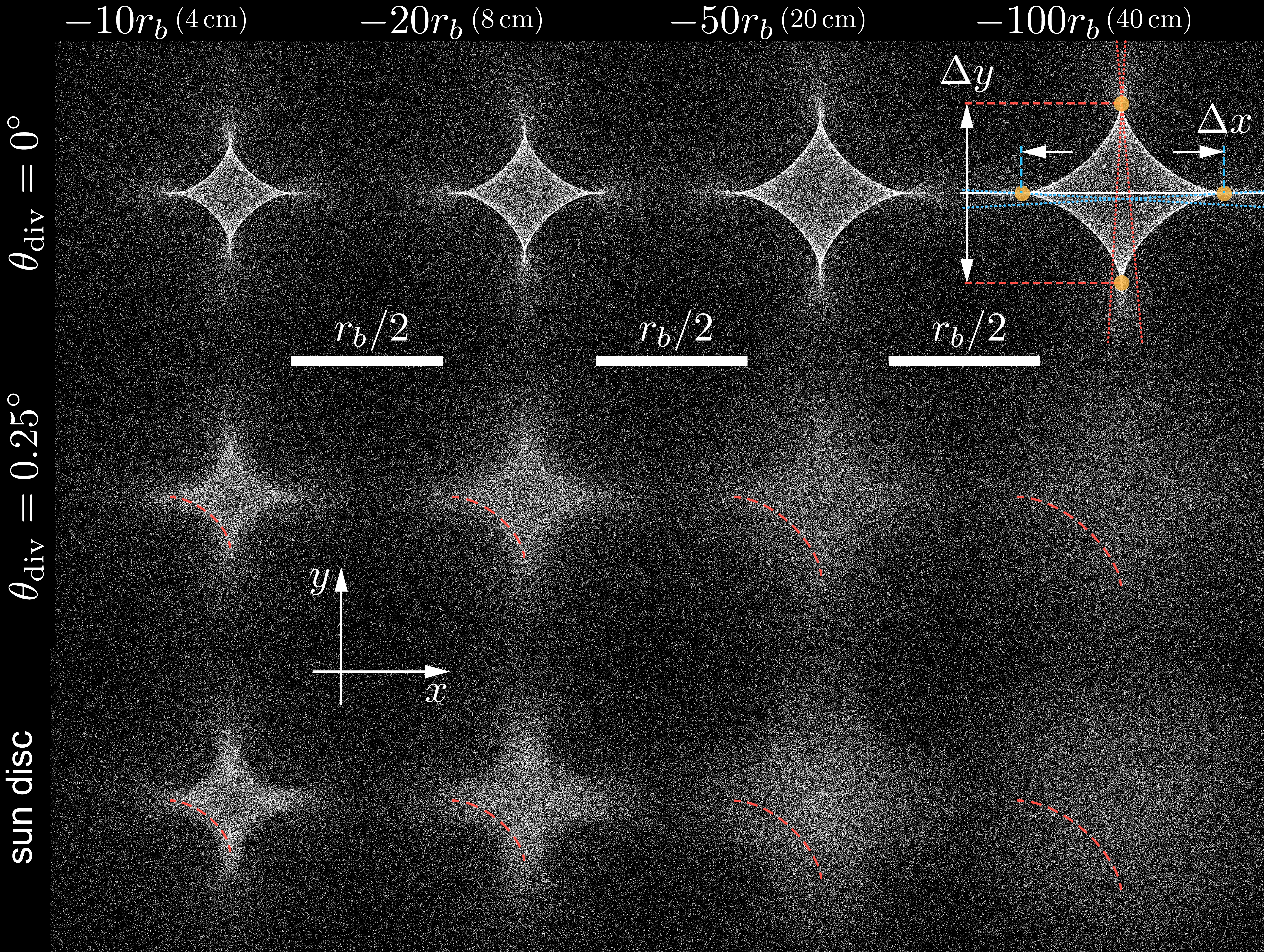}}
\caption{Ray tracings as in Fig.\ \ref{Fig_Formation} ($R_0/a=2$) but with divergence of light considered (cf.\ Fig.\ \ref{Fig_SketchEmergence}). The bubble caustics fade away with increasing distance (left to right) $z=z_p-z_\infty$ from the bubble (distances correspond to $z/(f_m-z_\infty)=(-1.3, -3.6, -6.6, -13)$). The first row is without light divergence, the second row with a Gaussian distribution with std $\sigma=0.25^{\circ}$, the third row assuming a solar disc of radius $0.25^{\circ}$. The solar disc caustics compare well with experiments such as those shown in Fig.\ \ref{Fig_Oblique}(g). The dotted lines (top right) show mapped radial spokes, and the orange discs the intersections used to determine the astroid size, cf.\ section \ref{sec:size} (the illustration is for a larger $\epsilon =\pi/50$).}%\lessapprox \gtrapprox $z_p=-10r_b, -20r_p, -50r_p, -100r_p = -4\rm \, cm, -8\rm \, cm, -20\rm \, cm, -40\rm \, cm$.
\label{Fig_Fade}
\end{figure}

Figure \ref{Fig_Fade} shows the effect of divergence for two scenarios: i) a Gaussian distribution of incidence ray angle perturbations $\phi_s$ around the average direction $\mathbf{k_1}$, and ii) a simulation using the approximation of a uniform solar disc (as an approximation to its "true" shape, and ignoring circumsolar radiation \cite{Buie2003}) of diameter $0.5^{\circ}$ around $\mathbf{k_1}$ for incidence ray directions (implemented by a beta distribution $\propto \phi_s$). Both scenarios may be thought of as a convolution of the astroid shape with a Gaussian or a disc filter. The effect in i) is a mere blurring of the astroid's shape into a more or less homogeneously lit star-like pattern with the four arm's ends remaining slightly augmented or brightened. The effect for scenario ii) is quite different: here, the astroid's shape morphs into a distinct \textit{cross-like shape} and an additional central dark x-shape appears for intermediate depths. For larger depths, the solar disc convolution with the astroid results in a box-like shape of the caustic pattern. Several of these features can be seen in Fig.\ \ref{Fig_Oblique}(g) for differently sized bubbles (effectively corresponding to different normalized depths). Qualitatively, one may expect the pattern to become indiscernible and the intensity to drop significantly when the size $\Delta y$ of the pattern at depth $|z|$ becomes comparable or smaller than the spreading distance of the solar rays travelling a distance $|z|$, i.e.\ $\Delta y/2 \lesssim \Delta \Gamma |z|$. For an $\varnothing 8\,\rm mm$-bubble ($R_0/a=2$) in water, and anticipating \eqref{eq:Fit}, this corresponds to depths larger than $\sim 0.2\rm\, m$. The observations noted in section \ref{sec:observations} agree with this order of magnitude estimate. %\lesssim

\subsection{The shadow sausage effect vs.\ the bubble caustic}\label{sec:sausage}
The bubble axicon caustic has incidentally already been computed explicitly in a certain limit in the context of the shadow-sausage effect. In this water-immersed stick caustic effect, only two cusps of the astroid caustic are visible and the other two are obstructed by the stick's shadow \cite{Lock2003}. Lock and coworkers analyzed the case of oblique light incidence on a vertical rod immersed in a liquid in their Section 5A, computing the astroid caustic in the gentle slope approximation. They had already drawn the connection to the relevant primary aberrations for this case: astigmatism and coma as they occur for skew rays through a spherical lens, referring to the aforementioned appendix 2 of Ref.\ [\!\!\citenum{Berry1980}] (in fact, it was their insightful remark which triggered the considerations in section \ref{sec:CatastropheTheory}). Since the exterior meniscus of a surface bubble is identical to a corresponding vertical rod, the astroid caustic of the bubble axicon may thus also be identified as a special case of an \textit{unobstructed shadow-sausage caustic} without the shadow. However, in contrast to the remark at the end of section 5A in \cite{Lock2003}, the \textit{inversion of the cusps} (i.e.\ the cross-over of the responsible rays) occurs \textit{in the direction along the incidence direction} (and not perpendicular to it). Experiments with the setup shown in Fig.\ \ref{Fig_Oblique}(a) and controlled illumination obstructions revealed (pictures not shown) this distinct behaviour and fine structure of the vertical rod caustic under inclined illumination as well as for the inclined rod caustic under vertical illumination. The more instructive ray tracing analysis (cf.\ section \ref{sec:RayTracing}), \textit{color-coded for each ray origin}, shows this more clearly and confirms the difference, see Fig.\ \ref{Fig_FalseColor}. A detailed parameter space analysis describing the transition between both cases is beyond the scope of this paper, but clearly the inversion behaviour of the vertical rod's shadow sausage effect caustic was hereby found to reconcile with the bubble axicon caustic. Also, the details of the plotted shadow sausage effect caustics suggests that the associated astroid involves a 4 to 2 ray transition, in contrast to the 3 to 1 ray transition hypothesized below loc.\ eq.\ (27) in Ref.\ [\!\!\citenum{Lock2003}]. This is at least the case for the unobstructed $\pm y$-caustics. The same holds true for the bubble caustic: While the four cusps each involve a 3 ray coalescence, there is 1 extra background ray making it 4 rays in total within the astroid, and 2 rays just outside of it reaching each point.

\subsection{Size of the astroid caustic}\label{sec:size}
As already noted by Lock et al.\ \cite{Lock2003} for the related limiting case of the vertical rod shadow sausage effect caustic, the lateral size $\Delta y$ of the bubble axicon astroid (see Fig.\ \ref{Fig_Fade}) \textit{grows with increasing inclination angle} $\Gamma$ of the incoming parallel light approximately as $\propto \sin^2(\Gamma)$. The effect can be seen for instance in the right-most column of Fig.\ \ref{Fig_Formation}. Moreover, the bubble axicon's astroid was found to \textit{grow with the distance} $z=z_p-z_\infty$ from the bubble, as can also be seen in Fig.\ \ref{Fig_Fade} going from left to right. For the purpose of this analysis, the size $\Delta y$ was determined from the intersection of mapped radial spokes $(r\cos(\phi),r\sin(\phi))$ with $r\in[r_b,3r_b]$ and $\phi=\pi/2\pm \epsilon$ and $\epsilon = \pi/200$, see Fig.\ \ref{Fig_Fade}. The lateral size $\Delta x$ was determined in a similar fashion, looking at the intersections of mapped radial spokes with $\phi=-\epsilon$ and $\phi=\pi+\epsilon$. Looking still exemplarily at the case $R_0/a=2$, and considering $\Delta y$ only ($\Delta x$ being enlarged again by the projection, cf.\ Fig.\ \ref{Fig_SketchEmergence}(a). For small bubbles with $R_0/a<1$, $\Delta x \approx \Delta y/\cos(\Gamma')$ was indeed found to hold.), the functional form
\begin{equation}
\Delta y/r_b \approx 2.56 \times \sin^2\left(0.00994\cdot \Gamma [^{\circ}]\right) \times |z/r_b|^{0.212}\label{eq:Fit}
%2.56 hTest^0.212 Sin[0.00994 \[CapitalGamma]Test]^2
\end{equation}
was found to fit the ray tracing data over a range of $\Gamma\in [0^{\circ},60^{\circ}]$ and $z\in[-10r_b,-150r_b]$ well (${<10\%}$ relative deviation for ${\Gamma>5^{\circ}}$ and all $h$, and at most $0.03$ absolute deviation for the entire parameter space spanning $\Delta y/r_b \in [0,2.1]$). For instance, the fit expression \eqref{eq:Fit} with $\Gamma=30^{\circ}$ and $z=-100r_b$ yields $\Delta y / r_b=0.59$, cf. also Fig.\ \ref{Fig_Fade} (top right).

Intuitively, the astroid's lateral size $\Delta y$ was also found to \textit{increase with the bubble size} $r_b$ (or the shape parameter $R_0/a$). For small bubbles with ${r_b/a<1}$, fitting of simulation data of $\Delta y/a$ for $R_0/a=(0.05, 0.1, 0.2, 0.5, 1, 2, 5, 10, 20, 50, 100, 300, 1000)$ and several fixed incidence angles $\Gamma=(10^{\circ}, 20^{\circ}, 30^{\circ})$ and fixed distances of $c\times f_m(R_0/a=0.05)$ and $c=(1.06,1.5)$ (i.e.\ for water at depths ${\sim 35\,\rm cm}$ and $50\,\rm cm$, respectively), showed an approximately linear dependence $\Delta y \propto r_b$ (or equivalently in $R_0/a$). For larger bubbles with ${r_b/a>1}$, the data revealed a power law scaling close to $\Delta y/a \propto (r_b/a)^{0.7}$.

As one would expect, ray tracings also showed that the astroid's size \textit{increases with decreasing distance $s$ of the point source} from infinity towards the bubble. For this, the incidence vectors instead of being fixed were set to $\mathbf{k_1}(x_i,y_i)=\mathbf{m_i}-s (-\sin(\Gamma),0,\cos(\Gamma))$ and each normalized to $|\mathbf{k_1}|=1$.

\subsection{Caustic observations}\label{sec:observations}
\begin{figure}[htbp]
\centering
\fbox{\includegraphics[width=\linewidth]{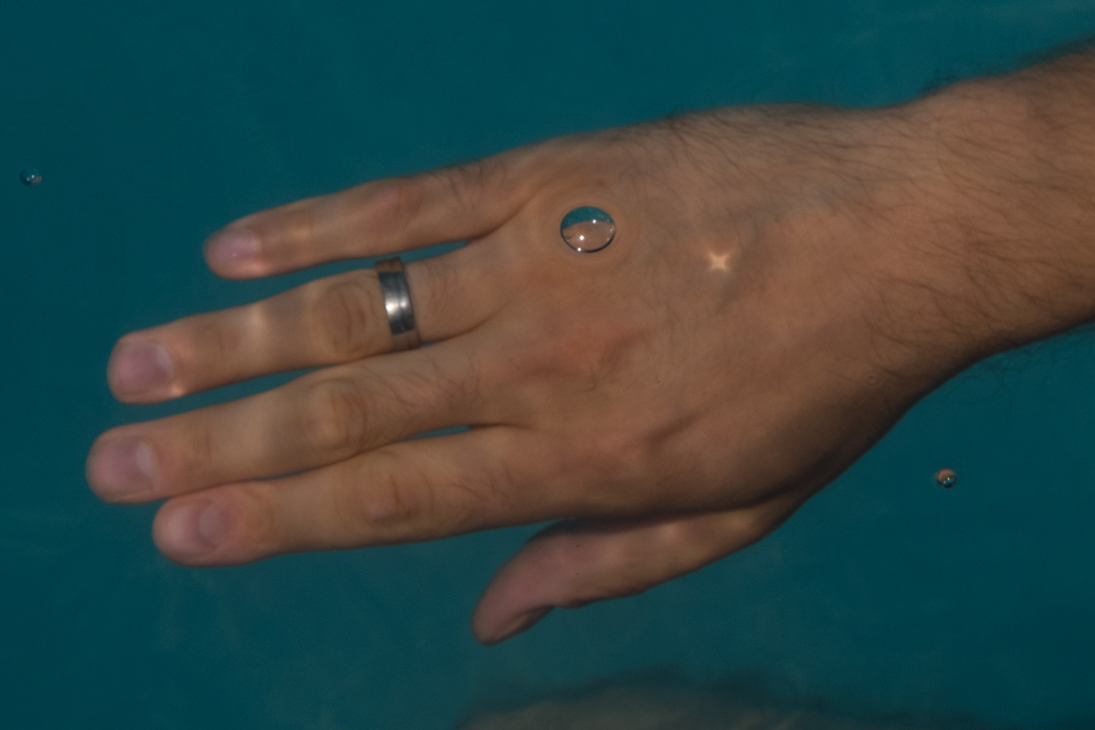}}
\caption{Photograph of a bubble axicon's astroid caustic for oblique light incidence (light elevation $e\sim 41^{\circ}$) as observed on a sunny day in a pool. Only below a critical depth of ${\sim 3 \,\rm cm}$, and only for afternoon or morning suns with low solar elevation (large $\Gamma$) are clear astroid caustics well visible. The hand could be moved way down (${\sim 1\,\rm m}$) without decreasing the perceived intensity much.}%\lessapprox \gtrapprox
\label{Fig_Pool}
\end{figure}

In accordance with the preceding analyses, the astroid caustic below individual obliquely illuminated bubbles can easily be observed, i.e.\ in a bathtub or an outdoor pool as photographed in Fig.\ \ref{Fig_Pool}. The caustics could be observed for bubbles of various sizes ($\lesssim 2\,\rm cm$) and and any depth $\gtrsim 3\, \rm cm$ of the pool ($\lesssim 2\,\rm m$, though becoming faint). Distortions due to close-by bubbles were hardly noticeable, provided they were separated by a few bubble radii. As expected, due to the angular divergence of sun light, the well-defined astroid caustics grew and eventually blurred to mere cross-like shapes with increasing depth before becoming only hardly discernible at the bottom of the pool (cf.\ Fig.\ \ref{Fig_Fade}, bottom row). Smaller bubbles created smaller astroids, and early morning and late afternoon observations generated the clearest and largest astroid caustics (large $\Gamma$), whereas at noon (near vertical incidence, smallest $\Gamma$) the foci appeared sharp (cf.\ Fig.\ \ref{Fig_Formation}). Aggregates of a few bubbles still acted as lenses, although the distortions altered the caustics visibly, especially in the near field. Collapsing bubbles produced dynamic concentric ring caustics, which in turn interacted with those by bubbles they encountered (not shown). The evolution of shadows to caustics could also be observed for larger bubbles, and agreed with the phenomenology shown in Fig.\ \ref{Fig_Formation}.

\begin{figure*}[htbp]
\centering
\fbox{\includegraphics[width=\linewidth]{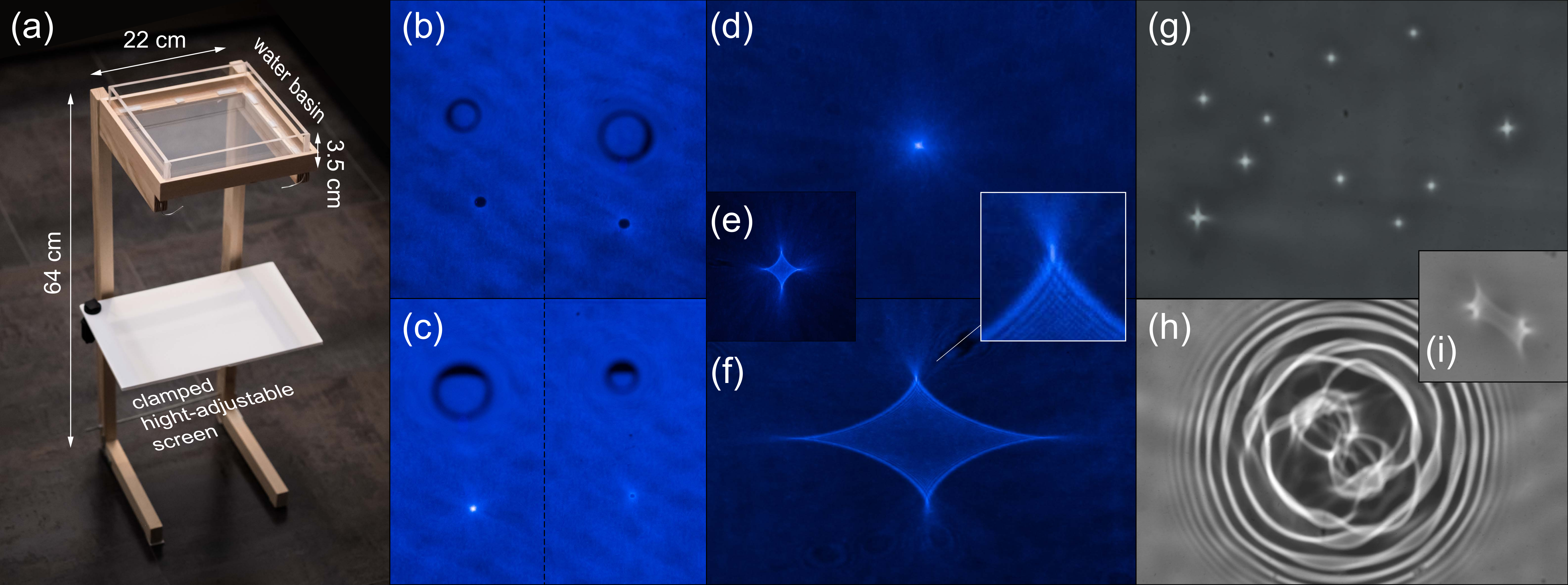}}
\caption{\textbf{(a)} Setup, consisting of a wooden mount, a screen (fixed with a camera equipment clamp) and an acrylic water basin (custom-cut acrylic $5\,\rm mm$-sheets, four pieces each $20\,{\rm cm}\times 3.5\,{\rm cm}$, one piece $20\,{\rm cm}\times 20\,{\rm cm}$, glued together using transparent aquarium silicone adhesive). The light was sunlight, or directed towards the setup via a ceiling-mounted mirror (camera equipment mounts) from a focusable laser diode ($405\,\rm nm$, $20\,\rm mW$) or a focusable LED flashlamp. Pictures were taken from (nearly) above (Fuji X-Pro 2, XF60mm F2.4 R Macro lens). \textbf{(b)} Shrunken bubble shadows surrounded by a bright halo. The water level was $d_t=1 \rm cm$, and the screen distance from the bottom of the basin was $0\rm\, cm$ and \textbf{(c)} $1\rm\, cm$. (small bubbles did not just yet develop the axial caustic) \textbf{(d)} Light source from above (elevation $e=90^{\circ}$), \textbf{(e)} Oblique illumination ($e\sim 20^{\circ}$), \textbf{(f)} and $e\sim 55^{\circ}$. The inset shows a magnification of the upper cusp with the characteristic Pearcey diffraction pattern. \textbf{(g)} Several differently sized bubbles giving well-isolated caustics under sunlight illumination ($e\sim 61^{\circ}$) and for a screen distance of about $30\,\rm cm$, showing a star-like pattern due to the solar rays' divergence, cf.\ Fig.\ \ref{Fig_Fade}. \textbf{(h)} Bubble coalescence event triggering the emission of capillary waves. \textbf{(i)} Two astroid caustics connect before a merger event.}%\lessapprox \gtrapprox
\label{Fig_Oblique}
\end{figure*}

A more controlled setup was also used to image shadows and caustics of bubbles (see Fig.\ \ref{Fig_Oblique}) and rods (cf.\ section \ref{sec:sausage}). Using laser light illumination, the bubble's four astroid cusps were confirmed to each being decorated by diffraction patterns as expected for these diffraction catastrophes \cite{Berry1980}. To this end, the laser beam was set to a highly oblique incidence angle, see Fig.\ \ref{Fig_Oblique}(f) and the inset: This allowed to capture the characteristic diffraction patterns photographically without further special equipment. Just like for the macroscopic shadow-sausage effect \cite{Lock2003}, no such patterns were observed by the unaided eye for sunlight or white-light illumination, probably because they were to fine \cite{Berry1994} or washed out by the light's divergence.

\section{Conclusion and Outlook}
Studying the refracting properties of floating bubbles, the outer meniscus was identified to act as a converging lens. Using the small bubble limit, analytical expressions were found which characterized the aberrations of such a lens. Via 2D ray tracing analysis, a shrinking shadow and behind it a semi-infinite focal range were found. Negative spherical aberration combined with positive refractive power provide the basis for the line caustic characteristic \cite{Burvall2004}. Eventually, the lens was identified as a special type of axicon lens. This was found to be consistent with the bubble's imaging and caustic characteristics. The unfolding of the naked axial caustic into a configuration of four fold caustics forming an astroid-shape, along with 3D ray tracing analysis, finally allowed for a detailed understanding of the observed bubble optics and its relation to the shadow-sausage effect and other natural axicons in astronomy and elsewhere.

Many further investigations of or using the caustics of bubble axicons are thinkable: \textit{First}, an increased flexibility and hence extension of the phenomenology is expected when considering a 3 phase system, i.e.\ a bubble of some liquid at an interface between two other liquids or a liquid and a gas \cite{Princen1965b}. The bubble axicon adds another lens variant, nearly perfect in shape by the action of smoothing surface tension forces, to the existing variable focus liquid lenses \cite{Stong1968,Chiu2012}. \textit{Second}, coalescing bubbles or doublets (multiplets) of bubbles in contact were found to yield interesting caustics, possibly adding a new model system useful for investigating (diffraction) catastrophes. This will be addressed in a separate study using the setup of Fig.\ \ref{Fig_Oblique}(a). \textit{Third}, the refractive interrogation of the bubble system as outlined in this paper could also be useful to study phenomena associated with bubble collapses: Fig.\ \ref{Fig_Axicon}(h) shows an image showing capillary waves emanating from coalesced and collapsed bubbles. Also directed ejection of smaller droplets from rupturing/bursting bubble caps \cite{Lhuissier2012,Bird2010} could be seen in slow-motion videos. \textit{Fourth}, coalescing bubbles could act as a toy system for gravitational lenses of binary mass systems (cf.\ Fig.\ \ref{Fig_Axicon}(i) and loc.\ Fig.\ 3.6 of Ref.\ [\!\!\citenum{Petters2001book}]). \textit{Fifth}, lithographic applications of the present system or variants thereof might be an interesting line of further research \cite{WeiLithography2008}. % (It has not escaped the author's attention that this would allow to extend meniscus studies like the ones presented here or elsewhere from floating particles to ''farticles''). 

Finally, it is the hope of the author that at any rate this article was successful at shedding some light on a fascinating everyday-phenomenon and providing some context for it. Maybe the reader will seek the astroid caustics, or think of the peculiar properties of axicons the next time he or she encounters a bubble in the bathtub, sink or outdoor pool.

\section{Acknowledgements}
I thank H.~Lhuissier for kindly providing his implementation of the numerical bubble shape profile algorithm \cite{Lhuissier2012}. I also thank M.\ V.\ Berry for helpful remarks, especially on the ray counts discussed in section \ref{sec:CatastropheTheory}.

%\clearpage

%\renewcommand{\arraystretch}{2}
%\renewcommand{\tabcolsep}{9pt}

%\bibliographystyle{unsrt}
%\bibliography{PhysTeacher}

\begin{thebibliography}{67}

\bibitem{Vella2005}
D.~Vella and L.~Mahadevana, ``The "Cheerios effect"", Am. J. Phys. \textbf{73}(9), 817--825 (2005).

\bibitem{Walker1983}
J.~Walker, ``Looking into the Ways of Water Striders, the Insects That Walk (and Run) on Water,'' Sci. Am. \textbf{249}(5), 188--197 (1983). %www.jstor.org/stable/24969039.

\bibitem{Adler1967}
C.~Adler, ``Shadow-Sausage Effect,'' Am. J. Phys. \textbf{35}(8), 774--776 (1967).

\bibitem{Walker1988}
J.~Walker, ``Shadows Cast on the Bottom of a Pool Are Not Like Other Shadows. Why?,'' Sci. Am. \textbf{259}(1), 116--119 (1988).

\bibitem{Lock2003}
J.A.~Lock, C.L.~Adler, D.~Ekelman, J.~Mulholland, and B.~Keating, ``Analysis of the Shadow-Sausage Effect Caustic,'' Appl. Optics \textbf{42}(3), 418--428 (2003).

\bibitem{Lock2015}
C.L.~Adler, J.A.~Lock, ``Caustics due to complex water menisci,'' Appl. Optics \textbf{54}(4), B207--B221 (2015).

\bibitem{Czerski2017}
H.~Czerski, ``Behold the bubbly ocean,'' Phys. World \textbf{30}(11), 34--38 (2017).

\bibitem{Seitz2011}
R.~Seitz, ``Bright water: hydrosols, water conservation and climate change,'' Clim. Change \textbf{105}(3-4), 365--381 (2011).

\bibitem{Davis1955}
G.E.~Davis, ``Scattering of Light by an Air Bubble in Water,'' J. Opt. Soc. Am. \textbf{45}(7), 572--581 (1955).

\bibitem{Marston1979}
P.L.~Marston, ``Critical angle scattering by a bubble: physical-optics approximation and observations,''  J. Opt. Soc. Am. \textbf{69}(9), 1205--1211 (1979).

\bibitem{Vera2001}
M.U.~Vera, A.~Saint-Jalmes, and D.J.~Durian, ``Scattering optics of foam,'' Appl. Optics \textbf{40}(24), 4210--4214 (2001).

\bibitem{Dyson1949}
J.~Dyson, ``Optical Diffraction Patterns Produced by Bubble Rafts,'' Proc. Roy. Soc. A \textbf{199}(1056), 130--139 (1949).

\bibitem{Hinsch1983}
K.~Hinsch, ``Holographic Study of Liquid Surface Deformations Produced by Floating Particles,'' J. Colloid Interface Sci. \textbf{92}(1), 243--255 (1983).

\bibitem{Wardle1970} %ellipsoid of revolution as approximation
M.W.~Wardle, H.J.~Gerritsen, ``Application of holographic interferometry to the static meniscus,'' Appl. Optics \textbf{9}(7), 1639--1642 (1970).

\bibitem{Mishra2015}
A.~Mishra, V.~Kulkarni, J.~Khor and S.T.~Wereley, ``Mapping surface tension induced meniscus with application to tensiometry and refractometry,'' Soft Matter \textbf{11}, 5619--5623, (2015).

\bibitem{DiLeonardo2003}
R.~Di Leonardo, F.~Saglimbeni, G.~Ruocco, ``Very-Long-Range Nature of Capillary Interactions in Liquid Films,'' Phys. Rev. Lett. \textbf{100}(10), 106103 (2008).

\bibitem{Hennequin2013}
Y.~Hennequin, C.P.~Allier, E.~McLeod, O.~Mudanyali, D.~Migliozzi, A.~Ozcan, J.-M.~Dinten, ``Optical Detection and Sizing of Single Nano-Particles Using Continuous Wetting Films,'' ACS Nano \textbf{7}(9), 7601--7609 (2013).

\bibitem{Berry1983}
M.V.~Berry, J.V.~Hajnal, ``The Shadows of Floating Objects and Dissipating Vortices,'' Opt. Acta \textbf{30}(1), 23--40 (1983).

\bibitem{Greenslade2012}
T.B.~Greenslade Jr., ``Surface Bubbles in the Bathtub and Reflections on Ripple Tanks,'' Phys. Teach. \textbf{50}(1), 17 (2012).% https://doi.org/10.1119/1.3670076

\bibitem{Shields1990}
J.~Shields, ``Swimming pool optics,'' Opt. Photonics News \textbf{1}(9), 37, 1990.

\bibitem{BashforthAdams1883}
F.~Bashforth and J.C.~Adams, \emph{An Attempt to Test the Theories of Capillary Action}
(University Press, Cambridge, 1883).

\bibitem{Nicolson1949}
M.N.~Nicolson, ``The interaction between floating particles,'' Math. Proc. Camb. Philos. Soc. \textbf{45}(2), 288--295, (1949).

\bibitem{Toba1959}
Y.~Toba, ``Drop Production by Bursting of Air Bubbles on the Sea Surface (II) Theoretical Study on the Shape of Floating Bubbles,'' J. Oceanogr. Soc. Jp. \textbf{15}(3), 1--10 (1959).

\bibitem{Chappelear1961}
D.C.~Chappelear, ``Models of a liquid drop approaching an interface,'' J. Colloid Sci. \textbf{16}, 186--190 (1961).

\bibitem{Princen1963}
H.M.~Princen, ``Shape of a fluid drop at a liquid-liquid interface,'' J. Colloid Sci. \textbf{18}, 178--195 (1963).

\bibitem{Princen1965a}
H.M.~Princen, S.G.~Mason, ``Shape of a fluid drop at a fluid-liquid interface - I. Extension and test of two-phase theory,'' J. Colloid Sci. \textbf{29}, 156--172 (1965).

\bibitem{Medrow1971}
R.A.~Medrow, B.T.~Chao, ``Floating Bubble Configurations,'' Phys. Fluids \textbf{14}(3), 459--465 (1971).

\bibitem{Teixeira2015}
M.A.C.~Teixeira, S.~Arscott, S.J.~Cox, P.I.C.~Teixeira, ``What is the Shape of an Air Bubble on a Liquid Surface?,'' Langmuir \textbf{31}, 13708--13717 (2015).

\bibitem{Puthenveettil2018}
B.A.~Puthenveettil, A.~Saha, S.~Krishnan, and E.J.~Hopfinger, ``Shape parameters of a floating bubble,'' Phys. Fluids \textbf{30}, 112105 (2018)

\bibitem{Lhuissier2012}
H.~Lhuissier, E.~Villermaux, ``Bursting bubble aerosols,'' J. Fluid Mech. \textbf{696}, 5--44 (2012)

\bibitem{Bird2010}
J.C.~Bird, R.~de Ruiter, L.~Courbin, H.A.~Stone, ``Daughter bubble cascades produced by folding of ruptured thin films,'' Nature \textbf{465}, 759--762 (2010).

\bibitem{Cohen2017}
C.~Cohen, B.D.~Texier, E.~Reyssat, J.H.~Snoeijer, D.~Qu\'er\'e, and C.~Clanet, ``On the shape of giant soap bubbles,'' PNAS \textbf{114}(10), 2515--2519 (2017).

\bibitem{Huh1969}
C.~Huh and L.E.~Scriven, ``Shapes of Axisymmetric Fluid Interfaces of Unbounded Extent,'' J. Colloid Interface Sci. \textbf{30}(3), 323--337 (1969).

\bibitem{White1965}
D.A.~White and J.A.~Tallmadge, ``Static menisci on the outside of cylinders,'' J. Fluid Mech. \textbf{23}(2), 325--335 (1965).

\bibitem{Tang2019}
Y.~Tang, S.~Cheng, ``The meniscus on the outside of a circular cylinder: From microscopic to macroscopic scales,'' J. Colloid Interface Sci. \textbf{533}, 401--408 (2019)

%======AXICONS==================
\bibitem{McLeod1954}
J.H.~McLeod, ``The Axicon: A New Type of Optical Element,'' J. Opt. Soc. Am. \textbf{44}(8), 592--597 (1954).

\bibitem{Rayces1958} %see also \cite{Vidals2017} p. 449
J.L.~Rayces, ``Formation of Axicon Images,'' J. Opt. Soc. Am. \textbf{48}(8), 576--578 (1958).

\bibitem{McLeod1960}
J.H.~McLeod, ``Axicons and their uses,'' J. Opt. Soc. Am. \textbf{50}(2), 166--169 (1960).

\bibitem{Zhang2013}
X.~Zhang, L.~Qiu, ``Generation of radially and azimuthally polarized light by achromatic meniscus axicon,'' Optical Engineering \textbf{52}(4), 048001 (2013).

\bibitem{Saikaley2013}
A.~Saikaley, B.~Chebbi, I.~Golub, ``Imaging properties of three refractive axicons,'' Appl. Optics \textbf{52}(28), 6910--6918 (2013).

\bibitem{Arimoto1992}
R.~Arimoto, C.~Saloma, T.~Tanaka, and S.~Kawata, ``Imaging properties of axicon in a scanning optical system,'' Appl. Optics \textbf{31}(31), 6653--6657 (1992) .

\bibitem{Burvall2004}
A.~Burvall, K.~Ko\l{}acz, Z.~Jaroszewicz, and A.T.~Friberg, ``Simple lens axicon,'' Appl. Optics \textbf{43}(25), 4838--4844 (2004).

\bibitem{Sochacki1992}
J.~Sochacki, A. Ko\l{}odziejczyk, Z. Jaroszewicz, S. Bar\'a, ``Nonparaxial design of generalized axicons,'' Appl. Optics \textbf{31}(25), 5326--5330 (1992).

\bibitem{Wang2017} %Cos-factor
Y.~Wang, S.~Yan, A.T.~Friberg, D.~Kuebel, and T.D.~Visser, ``Electromagnetic diffraction theory of refractive axicon lenses,'' J. Opt. Soc. Am. A \textbf{34}(7), 1201--1211 (2017) .

%=============================
\bibitem{Berry1980}
M.V.~Berry, C.~Upstill, ``Catastrophe optics: morphologies of caustics and their diffraction patterns,'' Prog. Opt. \textbf{18}, 257--346 (1980). 
%(ed. Wolf, E.) (North-Holland, Amsterdam, 1980).

\bibitem{Berry1981}
M.V.~Berry, ``Singularities in Waves and Rays,'' in Les Houches Lecture Series Session XXXV (Physics of Defects), eds. R. Balian, M. Kl\'eman and J.-P. Poirier, North-Holland: Amsterdam, 453--543 (1981). %1994?

\bibitem{Nye2005}
J.F.~Nye, ``The relation between the spherical aberration of a lens and the spun cusp diffraction catastrophe,'' J. Opt. A: Pure Appl. Opt. \textbf{7}, 95--102 (2005).

\bibitem{Berry1976}
M.V.~Berry, ``Waves and Thom's theorem,'' Adv. Phys. \textbf{25}(1), 1--26 (1976).

\bibitem{Nye1999Book}
J.F.~Nye, \emph{Natural Focusing and Fine Structure of Light: Caustics and Wave Dislocations}
(Institute of Physics Publishing, Philadelphia, 1999).

\bibitem{Nye1986} %p25 gives some examples of the defoc sequence of 2mm X9
J.F.~Nye, ``The Catastrophe Optics of Liquid Drop Lenses,'' Proc. R. Soc. Lond. A \textbf{403}, 1--26 (1986).

\bibitem{Nye2018}
J.F.~Nye, ``Symmetrical optical caustics,'' J. Opt. \textbf{20}, 075612 (2018).

\bibitem{Marston1984}
P.L.~Marston, E.H.~Trinh, ``Hyperbolic umbilic diffraction catastrophe and rainbow scattering from spheroidal drops,'' Nature \textbf{312}, 529--531 (1984).

\bibitem{Nye1984}%comment to Marston1984
J.F.~Nye, ``Rainbow scattering from spheroidal drops - An explanation of the hyperbolic umbilic foci,'' Nature \textbf{312}, 531--532 (1984).

\bibitem{Marston1989}
W.P.~Arnott, P.L.~Marston, ``Unfolding axial caustics of glory scattering with harmonic angular perturbations of toroidal wave fronts,'' J. Acoust. Soc. Am. \textbf{85}(4), 1427--1440 (1989).

\bibitem{Elliot1977}
J.L.~Elliot, R.G.~French, E.~Dunham, P.J.~Gierasch, J.~Veverka, C.~Church, ``Occultation of Epsilon Geminorum by Mars. II - The structure and extinction of the Martian upper atmosphere,'' Astrophys. J. \textbf{217}, 661--679 (1977).

\bibitem{Nicholson1995}
P.D.~Nicholson, C.A.~McGhee, R.G.~French, ``Saturn's central flash from the 3 July 1989 Occultation of 28 Sgr,'' Icarus \textbf{113}(1), 57--83 (1995).

\bibitem{Eshleman1979}
V.R.~Eshleman, G.L.~Tyler, W.T.~Freeman, ``Deep radio occultations and ''evolute flashes''; their characteristics and utility for planetary studies,'' Icarus \textbf{37}(3), 612--626 (1979).

\bibitem{Loutsenko2018}
I.~Loutsenko, ``On the role of caustics in solar gravitational lens imaging,''  Prog. Theor. Exp. Phys. \textbf{2018}(12), 123A02, 1--32, (2018).

\bibitem{Petters2001book} %coalesence: Fig. 3.6: caustics and critical curves due to 2 point masses with decreasing distance. See various elliptical gravitational lens models. See also Fig. 3.10
A.~O.~Petters, H.~Levine, J.~Wambsganss, \emph{Singularity Theory and Gravitational Lensing}
Progress in Mathematical Physics 21,
(Birkh\"auser Basel, Boston, 2001).

\bibitem{Surdej1993}
J.~Surdej, S.~Refsdal, and A.~Pospieszalska-Surdej, ``The Optical Gravitational Lens Experiment,'' in Gravitational Lenses in the Universe, Proceedings of the 31st Liege International Astrophysical Colloquium (LIAC 93), held June 21-25, 1993, edited by J.~Surdej, D.~Fraipont-Caro, E.~Gosset, S.~Refsdal, and M.~Remy (Institut d'Astrophysique, Liege, 1993), pp. 199--203. (I)

\bibitem{Lohre1996}
M.~Falbo-Kenkel, J.~Lohre, ``Simple gravitational lens demonstrations,'' , Phys. Teacher \textbf{34}(9), 555--557 (1996).

\bibitem{Buie2003} %http://spie.org/news/0975-bubble-and-antibubble-defects-in-193i-lithography
D.~Buie, A.G.~Monger, C.J.~Dey, ``Sunshape distributions for terrestrial solar simulations,'' Sol. Energy \textbf{74}(2), 113--122, (2003).

\bibitem{Berry1994}
M.V.~Berry, A.N.~Wilson ``Black-and-white fringes and the colours of caustics,'' Appl. Optics \textbf{33}, 4714--4718 (1994).

\bibitem{Princen1965b}
H.M.~Princen, S.G.~Mason, ``Shape of a fluid drop at a fluid-liquid interface - II. Theory for three-phase systems,'' J. Colloid Sci. \textbf{20}, 246--266 (1965).

\bibitem{Stong1968} %http://spie.org/news/0975-bubble-and-antibubble-defects-in-193i-lithography
C.L.~Stong, ``Experiments with Various Liquids That Do Not Mix,'' Sci. Am. \textbf{11}, (1968).

\bibitem{Chiu2012} 
C.-P.~Chiu , T.-J.~Chiang , J.-K.~Chen , F.-C.~Chang , F.- H.~Ko , C.-W.~Chu , S.-W.~Kuo, S.-K.~Fan, ``Liquid Lenses and Driving Mechanisms: A Review,'' J. Adhes. Sci. Technol. \textbf{26}(12-17), 1773--1788 (2012).

\bibitem{WeiLithography2008} %http://spie.org/news/0975-bubble-and-antibubble-defects-in-193i-lithography
Y.~Wei, ``Bubble and antibubble defects in 193i lithography, Air bubbles and topcoat particles act as microlenses, distorting patterns projected on the resist. Fourth in a series,'' SPIE Newsroom, 1-3, (2008).


\end{thebibliography}

\end{document}